\documentclass[epsfig,12pt]{article}
\usepackage{amssymb}
\usepackage{graphicx}
\usepackage{amsmath}

\input epsf.sty
\newtheorem{theorem}{Theorem}
\newtheorem{acknowledgement}[theorem]{Acknowledgement}

\input{tcilatex}
\makeatletter \@addtoreset{equation}{section}
\renewcommand{\theequation}{\thesection.\arabic{equation}}
\begin{document}

\title{\rightline{\mbox{\small
{Lab/UFR-HEP-0807-rev/GNPHE-0807-rev}}} \vspace{0.5cm} \textbf{Topological
String on Toric CY3s in Large Complex Structure Limit}}
\author{Lalla Btissam Drissi, Houda Jehjouh, El Hassan Saidi \\
\\
\vspace{-0.2cm} {\small 1. Lab/UFR- Physique des Hautes Energies, Facult\'{e}
des Sciences, Rabat, Morocco,}\\
{\small 2. GNPHE, Groupement National de Physique des Hautes Energies, Si%
\`{e}ge focal: FS, Rabat.}}
\maketitle

\begin{abstract}
We develop a non planar topological vertex formalism and we use it to study
the A-model partition function $\mathcal{Z}_{top}$ of topological string on
the class of toric Calabi-Yau threefolds (CY3) in large complex structure
limit. To that purpose, we first consider the $T^{2}\times R$ special
Lagrangian fibration of generic CY3-folds and we give the realization of the
class of large $\mu $ toric CY3-folds in terms of supersymmetric gauged
linear sigma model with \emph{non zero} gauge invariant superpotentials $%
\mathcal{W}\left( \Phi \right) $. Then, we focus on a one complex parameter
supersymmetric $U\left( 1\right) $ gauged model involving six chiral
superfields $\left\{ \Phi _{i}\right\} $ with $\mathcal{W}=\mu \left(
\prod\nolimits_{i=0}^{5}\Phi _{i}\right) $ and we use it to compute the
function $\mathcal{Z}_{top}$ for the case of the local elliptic curve in the
limit $\mu \rightarrow \infty $.\newline
{\small E-mails: \ \ drissilb@gmail.com, \ \ jehjouh@gmail.com, \ \
h-saidi@fsr.ac.ma}
\end{abstract}


\section{Introduction}

The\textrm{\ }discovery of topological tri-vertex formalism\footnote{%
In what follows, we shall refer to this formalism as topological \emph{planar%
} 3- vertex formalism.} by \emph{Aganagic et al} \textrm{\cite{vafa} } and
the developments that followed \textrm{\cite{refined}-\cite{tb}} have given
a big impulse to the study of topological string on local Calabi-Yau
threefolds $X_{3}$. Amongst the multiple results obtained in this direction,
we mention too particularly: (\textbf{i}) the explicit computation of the A
model topological string amplitudes for the set of toric Calabi-Yau
threefolds (CY3s) \textrm{\cite{vafa,t3}}. (\textbf{ii}) the works of Bryan
and Pandharipande who determined completely the local invariants of
nonsingular g-genus curves\ \textrm{\cite{31,32}}; and (\textbf{iii}) the
derivation of the generating function of Gromov--Witten invariants of toric
Calabi--Yau threefolds, which can be also expressed in terms of the
topological vertex \textrm{\cite{33}. }However, in most of these studies, a
special interest has been devoted to topological strings on those \emph{toric%
} CY3s which have a realization in terms of $2D$ $\mathcal{N}=2$
supersymmetric gauged linear sigma model without chiral matter
superpotential; that is $\mathcal{W}\left( \Phi \right) =0$. \newline
In this paper, we investigate the general situation where non zero chiral
superpotentials $\mathcal{W}\left( \Phi \right) \neq 0$ are implemented and
we study topological strings on that special class of \emph{toric}
Calabi-Yau threefolds $H_{3}$ describing backgrounds with $\mathcal{W}\left(
\Phi \right) \neq 0$.\newline
For this case, $\mathcal{W}\left( \Phi \right) \neq 0$, we will show amongst
others that the topological vertex formalism involves a \emph{non planar} (%
\emph{np}) topological vertex \QTR{cal}{C}$^{\left( np\right) }$ rather than
the standard planar one of ref \textrm{\cite{vafa}}. Moreover, considering
toric CY3s $H_{3}$ embedded in complex Kahler 4-folds $\mathcal{X}_{4}$; we
show equally that their topological vertex \QTR{cal}{C}$^{\left( np\right) }$
shares basic features of the topological 4- vertex associated with the
ambient space $\mathcal{X}_{4}$. The interpretation of \QTR{cal}{C}$^{\left(
np\right) }$ in terms of \textrm{3d- partitions }was exhibited in \textrm{%
\cite{DJS}}; but here we make a step further towards a formalism based on
2d-partitions by mainly using the decomposition property of 3d- partitions
(known as well as plane partitions) in terms of Young diagrams \textrm{\cite%
{sjd,AJ,34}}. \newline
Before going ahead, it is interesting to note that to deal with \QTR{cal}{C}$%
^{\left( np\right) }$ with rigor, sophisticated mathematical tools are
needed. Below, we will use rather a physical approach to shed more light on 
\QTR{cal}{C}$^{\left( np\right) }$ by taking advantage of the link between
Calabi-Yau manifolds and supersymmetric gauged linear sigma models. In this
optic, we first give \emph{useful} tools on toric geometry $\mathrm{\cite%
{s,s1,s3}}$ by focusing on the \emph{T}$^{2}\times R$ special Lagrangian
fibration of Calabi-Yau threefolds \textrm{\cite{jo,jds,Ri}} associated to
supersymmetric linear sigma models with \emph{non zero} gauge invariant
superpotentials $\mathcal{W}\left( \Phi \right) $. Then, we study\textrm{\ }%
the explicit expressions of the various hamiltonians of the $T^{2}\times R$
fibration of toric CY3s and we determine explicitly the values of the
shrinking cycles of the non planar vertices solving the Calabi-Yau
condition. This CY condition is physically interpreted in terms of the
conservation of total momenta at each vertex of the toric web diagram in the
same spirit as in the case of Feynman graph vertices of quantum field theory
(QFT)\textrm{. }\newline
As an illustration of the construction, we compute the A model topological
string partition function of the local elliptic curve \emph{in the} \emph{%
large complex structure limit} by using the cutting and gluing method gotten
by mimicking the \emph{Aganagic et al} approach.

The organization of this paper is as follows: In section 2, we briefly
describe the supersymmetric field theoretical set up of Calabi-Yau
threefolds with large complex structures. In section 3, we introduce helpful
tools for later use. We study the toric representations of local normal
bundle $NP^{n-1}$ for lower values of $n\leq 4$, by using supersymmetric
gauged linear sigma model. We also give useful results on these toric CY3s,
make comments on $T^{n-1}\times R$ special Lagrangian fibration and
determine the corresponding hamiltonians. In section 4, we study the
simplest example involving one gauge superfield $V$ and six chiral
superfields $\left\{ \Phi _{0},\Phi _{1},\Phi _{2},\Phi _{3},\Phi _{4},\Phi
_{5}\right\} $ with chiral superpotential $\mathcal{W}\left( \Phi \right) =%
\mathrm{\mu }\prod\nolimits_{i=0}^{5}\Phi _{i},$ $\mu \neq 0$. This gauge
invariant superfield model describes the local 2-torus in the large complex
structure limit. In section 5, we study the topological vertex formalism for
topological string on the local elliptic curve by using the non planar
vertex . In section 6, we give a conclusion and perspectives and in section
7, we give an appendix on local Gromov--Witten invariants of curves in a
Calabi--Yau threefold \textrm{\cite{t3,31}} and their \textrm{relationship }%
with the non planar topological vertex presented in this paper.

\section{\textbf{CY3s in large complex structure limit}}

To deal with the field theoretical set up of the complex deformations of
CY3s captured by gauge invariant chiral superpotential monomials, we start
by recalling the $2D$ $\mathcal{N}=2$ supersymmetric (or equivalently $4D$ $%
\mathcal{N}=1$) gauged linear sigma model with Lagrangian density $\mathrm{%
\cite{W}}$, 
\begin{equation}
\begin{tabular}{ll}
$L_{0}\left( x\right) =\int d^{2}\theta d^{2}\overline{\theta }\mathcal{L}%
\left( \Phi _{i},V_{a},\overline{\Phi }_{i};t^{a}\right) $ & ,%
\end{tabular}
\label{l}
\end{equation}%
describing the Kahler deformations of CY3s. In this relation, the Lagrangian
super- density $\mathcal{L}\left( \Phi _{i},V_{a},\overline{\Phi }%
_{i};t^{a}\right) $ is invariant under the following $U^{r}\left( 1\right) $
abelian gauge symmetry, 
\begin{equation}
\begin{tabular}{ll}
$\Phi _{i}^{\prime }=\left( e^{\mathrm{i}\left( \sum_{a=1}^{r}\mathrm{q}%
_{i}^{a}\Lambda _{a}\right) }\right) \Phi _{i},$ & , \\ 
$V_{a}^{\prime }=V_{a}-\mathrm{i}\left( \Lambda _{a}-\overline{\Lambda }%
_{a}\right) $ & ,%
\end{tabular}
\label{u}
\end{equation}%
with the $\mathrm{q}_{i}^{a}$ integers being the gauge charges of the matter
superfields and $\Lambda _{a}$ standing for chiral superfield gauge
parameters. The real superfields 
\begin{equation}
V^{a}\sim \left( A_{\mu }^{a},\text{ }\lambda _{\text{{\tiny Majorana}}}^{a},%
\text{ }D^{a}\right) ,
\end{equation}%
appearing in eq(\ref{l}) are $U^{r}\left( 1\right) $ abelian gauge
superfields; the $t^{a}$s are Kahler parameters and the hermitian super-
density $\mathcal{L}\left( \Phi _{i},V_{a},\overline{\Phi }_{i};t^{a}\right) 
$ describes the gauge invariant interacting dynamics of the $2D$ $\mathcal{N}%
=2$ chiral superfields $\Phi _{i}$ and $\overline{\Phi }_{i}$.

\emph{D-terms and CY3s\newline
}In this field theoretical formulation, the defining equation of local
Calabi-Yau threefolds $X_{3}$ is given by the usual equations of motion of
the auxiliary D-terms 
\begin{equation}
\begin{tabular}{ll}
$X_{3}:\qquad \frac{\delta L_{0}}{\delta D_{a}}=\sum_{i=1}^{3+r}q_{i}^{a}%
\left\vert \phi _{i}\right\vert ^{2}-t^{a}=0,\qquad a=1,...,r$ & ,%
\end{tabular}
\label{da}
\end{equation}%
where $\phi _{i}=\Phi _{i}$\TEXTsymbol{\vert}$_{\theta =0}$ are complex
scalar fields, often denoted as $z_{i}$, and where the real numbers $t^{a}$
are the Fayet-Iliopoulos (FI) real coupling constants. The real numbers $%
t^{a}$'s are interpreted geometrically as Kahler parameters capturing Kahler
deformations of the toric CY3 (\ref{da}). The Calabi-Yau condition,
requiring the vanishing of the first Chern class $\mathcal{C}_{1}\left(
X_{3}\right) =0$, translates in the superfield approach into the following
condition on the $U^{r}\left( 1\right) $ charges $q_{i}^{a}$ of the matter
superfields; 
\begin{equation}
\sum_{i=1}^{n}q_{i}^{a}=0,\qquad a=1,...,n-3,  \label{qi}
\end{equation}%
encoding the conformal behavior of the field theoretic model in the infrared 
\textrm{\cite{lwv,17,ABBDS}}.

Eqs(\ref{l}-\ref{qi}) are very well known in literature and they are not our
main purpose here; they are just tools towards the study of topological
strings on a special class of \emph{toric} Calabi-Yau threefolds $H_{3}$
going beyond the set of $X_{3}$'s described by eq(\ref{da}). Below, we shall
give details on the construction of the $H_{3}$s; but before notice that our
interest into this class of CY3s have been first motivated by looking for
the extension of the \emph{Aganagic et al} topological 3- vertex formalism
to the general case where the interacting gauge invariant dynamics of the
scalar fields $\phi _{i}$ contain, in addition to the usual \emph{%
gauge-matter} couplings namely, 
\begin{equation}
\int d^{4}\theta \sum_{a}\left( \sum_{i}q_{i}^{a}V_{a}\left\vert \Phi
^{i}\right\vert ^{2}-t^{a}V_{a}\right) ,
\end{equation}%
matter self- interactions captured by a non zero superpotential $\mathcal{W}%
\left( \Phi \right) $ $\mathrm{\cite{W}}$\textrm{-}$\mathrm{\cite{17}}$.

\emph{F- terms and the class of H}$_{3}$\emph{\ CY3s }\newline
In the case where there are matter self- interactions, $\mathcal{W}\left(
\Phi \right) \neq 0$, the above Lagrangian density $L_{0}$ extends as, 
\begin{equation}
L=L_{0}+\int d^{2}\theta \mathcal{W}\left( \Phi _{1},..,\Phi _{n};\mu
_{i}\right) +\int d^{2}\overline{\theta }\overline{\mathcal{W}}\left( 
\overline{\Phi }_{1},..,\overline{\Phi }_{n};\overline{\mu }_{i}\right) ,
\label{w}
\end{equation}%
where the complex coupling constants $\mu _{i}$ geometrically interpreted as
complex moduli of the underlying CY3. But gauge invariance of the
supersymmetric model severely restricts the family of the allowed polynomial
chiral superpotentials since, under the gauge change (\ref{u}), we should
have 
\begin{equation}
\begin{tabular}{ll}
$\mathcal{W}\left( \Phi _{1}^{\prime },...,\Phi _{n}^{\prime };\mu
_{i}\right) $ & $=\mathcal{W}\left( \Phi _{1},...,\Phi _{n};\mu _{i}\right) $%
, \\ 
$\overline{\mathcal{W}}\left( \overline{\Phi }_{1}^{\prime },...,\overline{%
\Phi }_{n}^{\prime };\overline{\mu }_{i}\right) $ & $=\overline{\mathcal{W}}%
\left( \overline{\Phi }_{1},...,\overline{\Phi }_{n};\overline{\mu }%
_{i}\right) $.%
\end{tabular}
\label{se}
\end{equation}%
Eqs(\ref{se}) put then a strong constraint on the allowed $\mathcal{W}\left(
\Phi \right) $s. A complex one parameter gauge invariant model for the
chiral superpotential $\mathcal{W}\left( \Phi \right) $ is given by the
typical monomial, 
\begin{equation}
\mathcal{W}\left( \Phi \right) \text{ }\sim \text{ }\mu
\prod\limits_{i=1}^{n}\Phi _{i},\qquad \overline{\mathcal{W}}\left( 
\overline{\Phi }\right) \text{ }\sim \text{ }\overline{\mu }%
\prod\limits_{i=1}^{n}\overline{\Phi }_{i},  \label{wo}
\end{equation}%
where $\mu $ is a complex coupling constant; describing a specific complex
modulus of the CY3. In the example (\ref{wo}), it is not difficult to see
that the constraint eqs(\ref{se}) are fulfilled due to the Calabi-Yau
condition eq(\ref{qi}). \newline
The supersymmetric equations of motion following from eq(\ref{w}) are given
by $\frac{\delta L}{\delta D_{a}}=\frac{\delta L_{0}}{\delta D_{a}}=0$ eq(%
\ref{da}); as well as the complex homomorphic ones, 
\begin{equation}
\begin{tabular}{ll}
$\frac{\delta \mathcal{W}}{\delta F_{i}}=0\qquad ,\qquad \frac{\delta 
\overline{\mathcal{W}}}{\delta \overline{F}_{i}}=0$ & ,%
\end{tabular}
\label{ft}
\end{equation}%
where $F$ and $\overline{F}$ are the well known auxiliary F-terms. We have%
\begin{equation}
\begin{tabular}{llll}
$\sum_{i=1}^{4+r}q_{i}^{a}\left\vert \phi _{i}\right\vert ^{2}-\zeta ^{a}=0$
& $,$ & $a=1,...,r$ & ,%
\end{tabular}
\label{ma}
\end{equation}%
together with%
\begin{equation}
\begin{tabular}{lllll}
$\mu \left( \prod\limits_{k=1,k\neq i}^{4+r}\phi _{k}\right) =0$ & $,$ & $%
\mu \left( \prod\limits_{k=1,k\neq i}^{4+r}\overline{\phi }_{k}\right) =0$ & 
$,i=1,..,4+r$ & $.$%
\end{tabular}
\label{mo}
\end{equation}%
From these relations, we learn that we should distinguish two main cases: 
\newline
\textbf{(a)} the case $\mu =0$, which corresponds to the usual
supersymmetric gauged linear sigma models. It mainly deals with Kahler
deformation moduli.\newline
\textbf{(b)} the case $\mu \neq 0$ describing a toric CY3- fold embedded in
a higher complex dimension Kahler manifold.\newline
Below, we will be interested by the large complex structure limit case 
\begin{equation}
\mu \mathrm{\qquad }\rightarrow \qquad \infty ,
\end{equation}%
so that (\ref{wo}) is thought of as the dominant term in the superpotential $%
\mathcal{W}\left( \Phi \right) $. A priori, a refined study involving more
than one complex parameter could be done without difficulty just by
implementing other gauge invariant monomials in $\mathcal{W}\left( \Phi
\right) $.

\emph{A field model for the local elliptic curve}\newline
To be more explicit, we consider hereafter one of the simplest
supersymmetric gauged model; namely the one involving the following degrees
of freedom:\newline
\textbf{(1)} One abelian gauge superfield $V:$ $r=1$.\newline
\textbf{(2)} Six chiral superfields $\Phi _{0},$ $\Phi _{1},$ $\Phi _{2},$ $%
\Phi _{3},$ $\Phi _{4},$ $\Phi _{5}$. For later use, we denote $\Phi _{5}$
as $\Upsilon $. \newline
\textbf{(3)} The total matter gauge invariant superpotential monomial, 
\begin{equation}
\mathcal{W}\left( \Phi \right) =\mu \left( \Upsilon
\prod\nolimits_{i=0}^{4}\Phi _{i}\right) ,\qquad \mu \neq 0.
\end{equation}%
The CY condition (\ref{qi}) for this model is solved as 
\begin{equation}
\left( q_{i}\right) =\left( -m,1,1,1,m,-3\right) ,
\end{equation}%
with $m$ an arbitrary integer which, for simplicity, we shall fix it to $m=3$%
. \newline
For the case $\mu =0$, the target space parameterized by the six complex
scalars $\phi _{i}$ describe the CY5- fold $\mathcal{O}(-m)\oplus \mathcal{O}%
(-3)\rightarrow WP_{1,1,1,m}^{3}$ where $WP_{1,1,1,m}^{3}$ stands for the
complex 3- dimension weighted projective space with the weights $\left(
1,1,1,m\right) $.\newline
For the case $\mu \neq 0$, the eqs of motions of the F-terms (\ref{ft}) are
non trivial since they capture extra constraints on the scalar fields as
shown below%
\begin{equation}
\begin{tabular}{llll}
$\frac{\partial W}{\partial \phi _{i}}=\mu \left( {\small \gamma }%
\prod\limits_{k\neq i}\phi _{k}\right) $ & $=0$ & $i=0,..,4$ & ,%
\end{tabular}
\label{gam}
\end{equation}%
and%
\begin{equation}
\begin{tabular}{lll}
$\frac{\partial W}{\partial \Upsilon }=\mu \left( \prod\limits_{k=0}^{4}\phi
_{k}\right) $ & $=0$ & .%
\end{tabular}
\label{gm}
\end{equation}%
In this case, the five eqs (\ref{gam}) can be collectively solved by taking $%
\gamma =0$ restricting the holomorphic set of field constraints to eq(\ref%
{gm}). Putting this value $\gamma =0$ back into eqs(\ref{mo}), we get%
\begin{eqnarray}
t &=&-m\left\vert \phi _{0}\right\vert ^{2}+\left\vert \phi _{1}\right\vert
^{2}+\left\vert \phi _{2}\right\vert ^{2}+\left\vert \phi _{3}\right\vert
^{2}+m\left\vert \phi _{4}\right\vert ^{2},  \label{a} \\
0 &=&\prod\limits_{k=0}^{4}\phi _{k},\qquad \mu \neq 0.  \label{b}
\end{eqnarray}%
The first relation of these eqs describe a complex 4 dimension Kahler sub-
manifold of the CY5- fold $\mathcal{O}(-m)\oplus \mathcal{O}(-3)\rightarrow
WP_{1,1,1,m}^{3}$. This sub- manifold is just $\mathcal{X}_{4}=\mathcal{O}%
(-m)\rightarrow WP_{1,1,1,m}^{3}$; but eq(\ref{a}) together with the
relation (\ref{b}) describe the Calabi-Yau threefold, 
\begin{equation}
H_{3}^{\left( m,-m,0\right) }=\mathcal{O}(m)\oplus \mathcal{O}%
(-m)\rightarrow E,  \label{1}
\end{equation}%
with $E$ being a complex curve given by three intersecting projective lines $%
E=P_{1}^{1}\cup P_{2}^{1}\cup P_{3}^{1}$ with matrix intersection $\mathcal{I%
}_{\alpha \beta }=P_{\alpha }^{1}\cap P_{\beta }^{1}$ as follows: 
\begin{equation}
\mathcal{I}_{\alpha \beta }=\left( 
\begin{array}{ccc}
-2 & 1 & 1 \\ 
1 & -2 & 1 \\ 
1 & 1 & -2%
\end{array}%
\right) ,\qquad \alpha ,\beta =1,2,3.  \label{mi}
\end{equation}%
In the language of toric geometry where the projective lines $P^{1}$ are
described by segments and the projective plane $P^{2}$ is represented by a
triangle $\Delta $, the complex curve $E$ can be imagined as the toric
boundary $\left( \partial P^{2}\right) $ of the complex projective plane $%
P^{2}$. It can be thought as well as the toric realization of the elliptic
curve in the large complex structure limit; $\mathrm{\mu }\rightarrow \infty 
$. The toric web diagram of E is given by the boundary of the triangle $%
\left( \partial \Delta \right) $. \newline
The toric Calabi-Yau threefold (\ref{1}) is then recovered by gluing the
toric threefolds 
\begin{equation*}
Y_{\alpha }=\mathcal{O}(m)\oplus \mathcal{O}(-m)\rightarrow P_{\alpha
}^{1},\alpha =1,2,3,
\end{equation*}%
with matrix intersection $\mathcal{I}_{\alpha \beta }$ in the base manifold
as in eq(\ref{mi}). As we will see later, the non planar topological vertex 
\QTR{cal}{C}$^{\left( np\right) }$ associated with the toric $H_{3}^{\left(
m,-m,0\right) }$ turns out to be given by the fusion of at least two planar
topological vertices belonging to different planes; more details are
exhibited in sections 4 and 5.

\section{Toric representations of local $NP^{n-1}$}

In this section we review useful aspects of the $NP^{n-1}$ toric model for
lower values of $n$, namely $n=2,3$ and $4$. The Calabi-Yau manifolds $%
NP^{n-1}$ are the simplest toric Calabi-Yau varieties on which we can
illustrate most of the basic geometric properties of local Calabi-Yau
manifolds. But before going ahead, let us recall briefly the field content
and the superfield action of $NP^{n-1}$ for $n\geq 2$.\newline
In the $4D$ $\mathcal{N}=1$ superfield set up which is equivalent to $2D$ $%
\mathcal{N}=2$ formalism, the complex $n$ dimension local manifold 
\begin{equation*}
NP^{n-1}\equiv \mathcal{O}\left( -n\right) \rightarrow P^{n-1},
\end{equation*}%
is the target space of the $U\left( 1\right) $ gauged supersymmetric linear
sigma model consisting of:\newline
(1) $\left( n+1\right) $ chiral superfields $\left\{ \Phi _{0},\Phi
_{1},\ldots ,\Phi _{n}\right\} $ carrying the $U\left( 1\right) $\ charges$\
\left( -n,1,\ldots ,1\right) $ that satisfy the Calabi-Yau condition\textrm{%
\ }$\sum_{i=0}^{n}q_{i}=0$.\newline
(2) An abelian $U\left( 1\right) $ gauge superfield $V$ which reads, in
terms of the $4D$ $\mathcal{N}=1$ superspace coordinates $\left( x,\theta ,%
\overline{\theta }\right) $ and the component fields $\left( A_{\mu
},\lambda ^{a},\overline{\lambda }_{\dot{a}},D\right) $, as follows, 
\begin{equation}
V\left( x,\theta ,\overline{\theta }\right) =-\theta \sigma ^{\mu }\overline{%
\theta }A_{\mu }-i\overline{\theta }^{2}\theta \lambda +i\theta ^{2}%
\overline{\theta }\overline{\lambda }+\frac{1}{2}\theta ^{2}\overline{\theta 
}^{2}D.  \label{bsi1}
\end{equation}%
The associated superfield Lagrangian density is 
\begin{equation}
L_{NP^{n-1}}=\int d^{4}\theta \left( \sum_{i=0}^{n}\overline{\Phi }%
_{i}e^{2q_{i}V}\Phi _{i}-2tV\right) +L_{gauge}\left( A_{\mu },\lambda ^{a},%
\overline{\lambda }_{\dot{a}},D\right) .  \label{bsi5}
\end{equation}%
Notice that the equation of motion of the auxiliary D-field, namely $\left(
\partial L_{NP^{n-1}}/\partial D\right) =0,$ leads to the defining equation
of $O\left( -n\right) \rightarrow P^{n-1}$: 
\begin{equation}
-n\left\vert z_{0}\right\vert ^{2}+\sum_{i=1}^{n}\left\vert z_{i}\right\vert
^{2}=t.  \label{bsi6}
\end{equation}%
In the following we focus our interest on the leading $n=2,3,4$ local
manifolds $NP^{n-1}$ that are relevant for the study of: \newline
\textbf{(i)} Low energy supersymmetric effective field theory limit of \emph{%
10D} type IIA superstring compactification down to lower dimensions, in
particular to four space time dimensions.\newline
\textbf{(ii)} Topological string on Calabi-Yau threefolds which is a
powerful method to deal with the type II\ superstring perturbation theory.
The fact that these $NP^{n-1}$ manifolds are toric is an important property
for the use of the \emph{Aganagic et al} topological vertex method for
computing topological string amplitudes.

\subsection{$NP^{1}$ as a linear geometry}

We first study the $T^{2}$\ fibration of $NP^{1}$ and then we consider its $%
S^{1}\times R$\ special fibration. This analysis should be understood as an
illustration of the main idea.

\subsubsection{$T^{2}$\textit{\ fibration of }$NP^{1}$}

First of all, notice that algebraically, the toric diagram describing the
complex projective line $P^{1}$ is given by the dimension 1-simplex $B_{1}$
fibered by a circle $S^{1}$. The base $B_{1}$ is a segment in the real plane
as shown below: 
\begin{equation}
B_{1}=\left\{ \left( x_{1},x_{2}\right) \in R_{+}^{2}\text{ }|\text{ }%
x_{1}+x_{2}=t\right\} ,
\end{equation}%
On each point $\left\{ x\right\} $ of the base $B_{1}$ lives a real 1-cycle $%
S_{x}^{1}$ so that the projective line $P^{1}$ can be thought of as 
\begin{equation}
P^{1}\text{ }\sim \text{ }\dcoprod\limits_{x\in B_{1}}\left( \left\{
x\right\} \times S_{x}^{1}\right) .
\end{equation}%
The $S_{x}^{1}$ fiber shrinks to zero on the two base's ends $\left(
x_{1},x_{2}\right) =\left( t,0\right) $ and $\left( 0,t\right) $.
Geometrically, the base $B_{1}$ is a finite straight line in the $\frac{1}{4}
$- plane $\left( x_{1},x_{2}\right) $ since $x_{1},$ $x_{2}\geq 0$. In the
limit $t$ goes to zero, we have 
\begin{equation}
x_{1},\text{ }x_{2}\longrightarrow 0
\end{equation}%
and the 1-simplex $B_{1}$ shrinks to the origin of the plane where lives an $%
SU\left( 2\right) $ singularity described by the ALE space $NP^{1}$.\newline
\begin{figure}[tbph]
\begin{center}
\hspace{-1cm} \includegraphics[width=10cm]{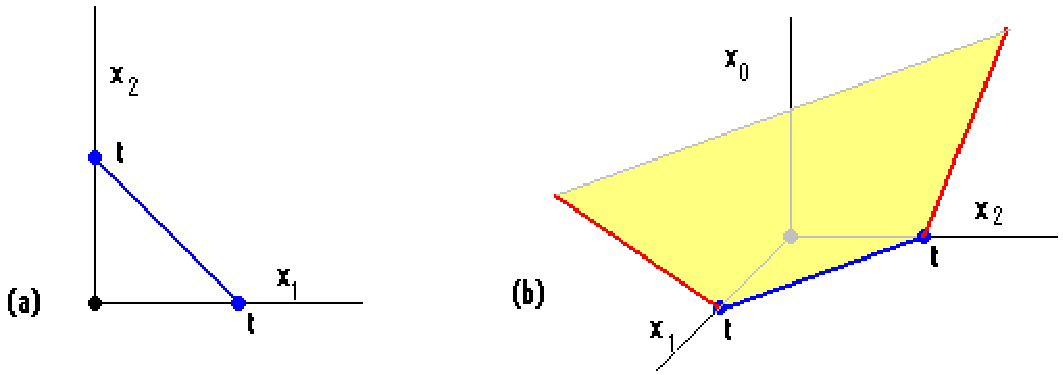}
\end{center}
\caption{{\protect\small (\textbf{a}) Toric diagram of }$P^{1},$%
{\protect\small \ (\textbf{b}) Toric diagram of }$\mathcal{O}\left(
-2\right) \rightarrow P^{1}$}
\end{figure}
Recall also that the local space $NP^{1}$ is a toric complex surface
capturing a natural $T^{2}$ fiber parameterized by the phases $\vartheta _{i}
$ of the three complex variables $z_{0}$, $z_{1}$ and $z_{2}$, moded out by
the $U\left( 1\right) $ gauge symmetry of eq(\ref{bsi6}). The corresponding
toric graph is given by a non compact real surface $B_{2}$, 
\begin{equation}
x_{0}=\frac{1}{2}\left( x_{1}+x_{2}-t\right) \text{ },\qquad x_{i}\geq 0%
\text{ \ },  \label{bsi8}
\end{equation}%
with a $T^{2}$ fiber.We have 
\begin{equation}
NP^{1}=\cup _{p\in B_{2}}\left( \left\{ p\right\} \times T_{p}^{2}\right)
,\qquad p=\left( x,y\right) ,  \label{T2}
\end{equation}%
which we we denote formally as $T^{2}\times B_{2}$. Note that $NP^{1}$ is a
toric surface, its boundary $\partial \left( NP^{1}\right) $ is also toric
and is given by a $S_{x}^{1}$ fibration over the boundary real line 
\begin{equation}
L=L_{1}\cup L_{2}\cup L_{0}
\end{equation}%
with 
\begin{equation}
\left\{ 
\begin{array}{c}
L_{1}:x_{2}-2x_{0}-t=0,\qquad x_{1}=0 \\ 
L_{2}:x_{1}-2x_{0}-t=0,\qquad x_{2}=0 \\ 
L_{0}:x_{1}+x_{2}-t=0,\qquad x_{0}=0%
\end{array}%
\right. .  \label{bsi9}
\end{equation}%
The corresponding toric diagrams of eqs(\ref{bsi8}-\ref{bsi9}) are reported
in \emph{figure}\textrm{\ }(1a)-(1b). \newline
For later use, it is interesting to consider the $S^{1}\times R$ fibration
of $NP^{1}$ where the previous $T^{2}=S^{1}\times S^{1}$ fiber eq(\ref{T2})
gets replaced by $S^{1}\times R$. The second $S^{1}$\ of $T^{2}$ has been
decompactified to $R$. More details are given below.

\subsubsection{$S^{1}\times R$ \textit{fibration of }$NP^{1}$}

In this setting, the local surface $NP^{1}$, and in general any smooth
Calabi-Yau 2-fold, can be obtained by gluing together $C^{2}$ patches in a
way that is consistent with Ricci-flatness. The geometry of $NP^{1}$ (and
any Calabi-Yau 2-fold) is encoded in the one dimensional graph $\Gamma _{1}$%
\ in the base that corresponds to the degeneration locus of the fibration.
In the example of $NP^{1}$, shown on \emph{figure 2}, we have two (bivalent)
vertices $V_{1}$ and $V_{2}$. The edges $E_{1},$ $E_{2}$ and $E_{3}$ of the
graph $\Gamma _{1}$ are oriented straight lines labeled by integers $%
p_{i}\in Z^{\ast }$ describing the shrinking 1-cycle $a_{i}\in H_{1}\left(
S^{1}\right) $. 
\begin{figure}[tbph]
\begin{center}
\hspace{-1cm} \includegraphics[width=10cm]{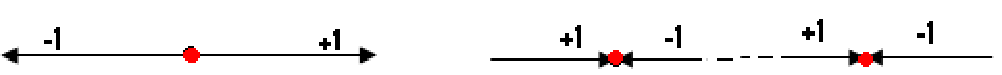}
\end{center}
\caption{{\protect\small The graph of }$S^{1}\times R${\protect\small \
fibration of }$O\left( -2\right) \rightarrow P^{1}${\protect\small . The
local complex surface is built out of two }$C^{2}${\protect\small \ patches
with orientations as indicated in the figure. The transition functions
correspond to a Z}$_{2}${\protect\small \ transformation of the }$S^{1}$%
{\protect\small \ fibers as one goes from one patch to the other.}}
\end{figure}
\newline
The condition of being a smooth Calabi-Yau is equivalent to the condition
that on each vertex $V$, if we choose the edges to be outgoing with charges $%
p_{i}$, we must have 
\begin{equation}
p_{1}+p_{2}=0.
\end{equation}%
For the case at hand, we have 
\begin{equation}
\left\{ 
\begin{array}{c}
\text{vertex }V_{1}:\text{ }p_{11}+p_{12}=0 \\ 
\text{vertex }V_{2}:\text{ }p_{21}+p_{22}=0%
\end{array}%
\right. .  \label{bsi10}
\end{equation}%
Changing the orientation on each edge $E_{i}$ corresponds to replacing $%
p_{i}\rightarrow -p_{i}$ which does not change the Calabi-Yau geometry. The
graph of the $S^{1}\times R$ \textit{fibration }of the local surface $NP^{1}$%
, which involves two open sets 
\begin{equation}
U_{1}\left( z_{0},z_{1}\right) \simeq C^{2},\qquad U_{2}\left(
z_{0},z_{2}\right) \simeq C^{2},
\end{equation}%
can be obtained as follows:\newline
Let $z_{i}$ be local complex coordinates on $C^{2}$, $i=0,1$. In the patch $%
U_{1}\left( z_{0},z_{1}\right) $, the base of the $S^{1}$ fibration is the
image of the moment map 
\begin{equation}
r_{\alpha }=\left\vert z_{1}\right\vert ^{2}-\left\vert z_{0}\right\vert
^{2},
\end{equation}%
which reads in the patch $U_{2}\left( z_{0},z_{2}\right) $ as 
\begin{equation*}
r_{\alpha }=-\left\vert z_{2}\right\vert ^{2}\ +\left\vert z_{0}\right\vert
^{2}+t.
\end{equation*}%
The non compact direction R is generated by $r_{\beta }=\func{Im}\left(
z_{0}z_{1}z_{2}\right) $. The special Lagrangian fiber is then generated by
the action of the two \textquotedblleft Hamiltonians\textquotedblright\ $%
r_{\alpha }$ and $r_{\beta }$ on $C^{2}$ via the standard symplectic form $%
\omega =i\left( dz_{0}\wedge d\overline{z}_{0}+dz_{1}\wedge d\overline{z}%
_{1}\right) $ on $C^{2}$ and the Poisson brackets $\partial z_{i}=\left\{
r_{\alpha },z_{i}\right\} _{\omega }$. Note that the $S^{1}$ fiber is
generated by the $U\left( 1\right) $ action 
\begin{equation}
e^{i\alpha r_{\alpha }}:\left( z_{0},z_{1}\right) \rightarrow \left(
e^{-i\alpha }z_{0},e^{i\alpha }z_{1}\right) ,  \label{bsi11}
\end{equation}%
which degenerates over $z_{0}=0=z_{1}$.

\subsection{$NP^{2}$ as a planar geometry}

\subsubsection{$T^{3}$\textit{\ fibration of }$\mathcal{O}\left( -3\right)
\rightarrow \mathbb{P}^{2}$}

The toric diagram describing the complex projective surface $P^{2}$ is given
by $T^{2}$\textit{\ }fibration over\textit{\ }the dimension 2-simplex 
\begin{equation}
B_{2}:x_{1}+x_{2}+x_{3}=t,
\end{equation}%
where $t$ is the $P^{2}$ Kahler parameter. $B_{2}$ is a finite equilateral
triangle embedded in the $\mathbb{R}_{+}^{3}$ octant $\left(
x_{1},x_{2},x_{3}\right) $. In the limit $t$ goes to zero, $B_{2}$ shrinks
to the origin of the octant where lives a $P^{2}$ singularity. 
\begin{figure}[tbph]
\begin{center}
\hspace{-1cm} \includegraphics[width=10cm]{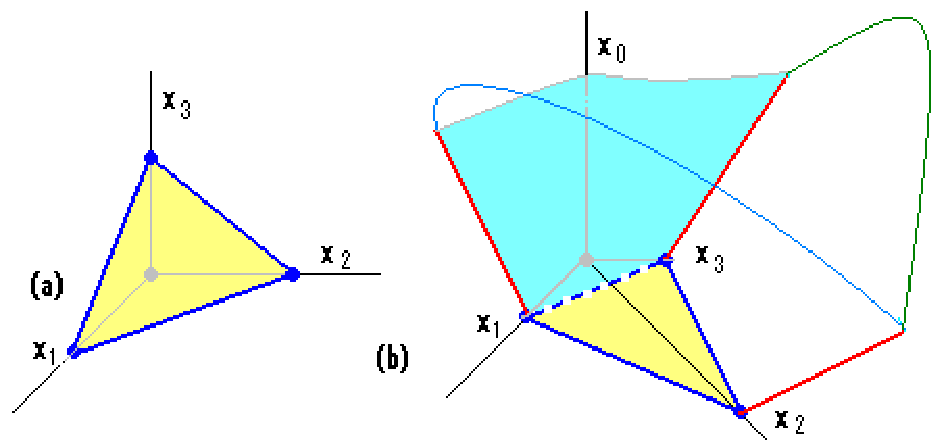}
\end{center}
\caption{{\protect\small (\textbf{a}) Toric diagram of }${\protect\small P}%
^{2}${\protect\small . (b) Toric diagram of }$\mathcal{O}\left( -3\right)
\rightarrow P^{2}${\protect\small \ where divisors have been also
represented.}}
\label{01}
\end{figure}
In the case of the normal bundle $\mathcal{O}\left( -3\right) \rightarrow
P^{2}$ viewed as a $T^{3}$ fibration over a real base $B_{3}$, the
corresponding toric graph is given by, 
\begin{equation}
x_{0}=\frac{1}{3}\left( x_{1}+x_{2}+x_{3}-t\right) ,\qquad x_{i}\geq 0.
\label{bsi13}
\end{equation}%
with a $T^{3}$ fiber; i.e 
\begin{equation}
NP^{2}\sim T^{3}\times B_{3}.
\end{equation}%
Note that $NP^{2}$ is toric and its boundary (divisor) is toric given by a $%
T^{2}$ fibration over the real surface 
\begin{equation}
D=D_{0}\cup D_{1}\cup D_{3}\cup D_{4},
\end{equation}%
with, 
\begin{equation}
\left\{ 
\begin{array}{c}
D_{0}:x_{1}+x_{2}+x_{3}-t=0,\qquad x_{0}=0 \\ 
D_{1}:x_{2}+x_{3}-3x_{0}-t=0,\qquad x_{1}=0 \\ 
D_{2}:x_{1}+x_{3}-3x_{0}-t=0,\qquad x_{2}=0 \\ 
D_{3}:x_{1}+x_{2}-3x_{0}-t=0,\qquad x_{3}=0%
\end{array}%
\right.  \label{bsi14}
\end{equation}%
The corresponding diagrams of eqs(\ref{bsi13}-\ref{bsi14}) are reported in 
\emph{figure} (2a) and (2b).

\subsubsection{$T^{2}\times R$ \textit{fibration of }$NP^{2}$}

The local surface $NP^{2}$, and in general any smooth toric Calabi-Yau
3-fold, can be obtained by gluing together $C^{3}$ patches in a way that is
consistent with Ricci-flatness. The geometry of $NP^{2}$ (and any Calabi-Yau
3-fold) is encoded in a planar graph $\Gamma _{2}$\ in the base that
corresponds to the degeneration locus of the fibration. In the present
example, shown on \emph{figure 3}, we have three (trivalent) vertices $V_{1}$%
, $V_{2}$ and $V_{3}$. The edges $E_{i}\ $of the graph $\Gamma _{2}$ are
oriented straight lines labeled by integer 2-vectors 
\begin{equation}
\mathbf{v}_{i}=\left( p_{i},q_{i}\right) \in Z^{2}
\end{equation}%
describing the shrinking 1-cycle $a_{i}\in H_{1}\left( T^{2}\right) $. 
\begin{figure}[tbph]
\begin{center}
\hspace{-1cm} \includegraphics[width=8cm]{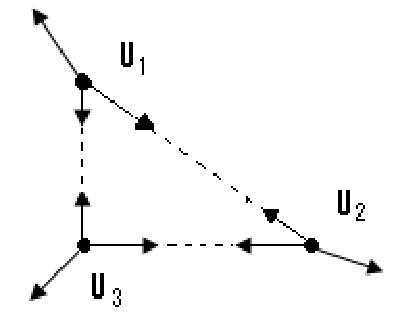}
\end{center}
\caption{{\protect\small The graph of }$T^{2}\times R${\protect\small \
fibration of }$O\left( -3\right) \rightarrow P^{2}${\protect\small . It is
built out of three }$C^{3}${\protect\small \ patches as indicated in the
figure.}}
\end{figure}
\newline
The condition of being a smooth toric Calabi-Yau is equivalent to the
condition that on each vertex $V_{i}$, if we choose the edges to be outgoing
with charges $\mathbf{v}_{i}$, we must have \ 
\begin{equation}
\sum_{i=1}^{3}\mathbf{v}_{i}=0.
\end{equation}%
For the case at hand, we have 
\begin{equation}
\left\{ 
\begin{array}{c}
\text{vertex }V_{1}:\text{ }\mathbf{v}_{11}+\mathbf{v}_{12}+\mathbf{v}_{13}=0
\\ 
\text{vertex }V_{2}:\text{ }\mathbf{v}_{21}+\mathbf{v}_{22}+\mathbf{v}_{23}=0
\\ 
\text{vertex }V_{3}:\text{ }\mathbf{v}_{31}+\mathbf{v}_{32}+\mathbf{v}_{33}=0%
\end{array}%
\right. .  \label{bsi15}
\end{equation}%
Changing the orientation on each edge $E_{i}$ corresponds to replacing $%
\mathbf{v}_{i}\rightarrow -\mathbf{v}_{i}$ which does not change the
Calabi-Yau geometry. The graph of the $T^{2}\times \mathbb{R}$ \textit{%
fibration }of the\textit{\ }local surface $NP^{2}$ involves three patches 
\begin{equation}
\begin{tabular}{llllll}
$U_{1}\left( z_{0},z_{2},z_{3}\right) \simeq $ $C^{3}$ & $,$ & $U_{2}\left(
z_{0},z_{1},z_{3}\right) \simeq $ $C^{3}$ & $,$ & $U_{3}\left(
z_{0},z_{1},z_{2}\right) \simeq $ $C^{3}$ & ,%
\end{tabular}%
\end{equation}%
\ and can be obtained as follows:\newline
In the local patch $U_{3}\left( z_{0},z_{1},z_{2}\right) $, the base of the $%
T^{2}$ fibration is the image of moment maps 
\begin{equation}
\left\{ 
\begin{array}{c}
r_{\alpha }=-\left\vert z_{0}\right\vert ^{2}+\left\vert z_{1}\right\vert
^{2} \\ 
r_{\beta }=-\left\vert z_{0}\right\vert ^{2}+\left\vert z_{2}\right\vert ^{2}%
\end{array}%
\right. .  \label{bsi16}
\end{equation}%
From these relations, one can read the $U\left( 1\right) \times U\left(
1\right) $ charges $\left( \alpha ,\beta \right) $ of the $z_{i}$ variables.
We have for $z_{0}$, $z_{1}$ and $z_{2}$ respectively 
\begin{equation}
\begin{tabular}{lllll}
$\mathbf{v}_{11}=\left( -1,-1\right) $ & $,$ & $\mathbf{v}_{12}=\left(
1,0\right) $ & , & $\mathbf{v}_{13}=\left( 0,1\right) .$%
\end{tabular}
\label{bsi17}
\end{equation}%
Similarly, we write down the $r_{\alpha }$ and $r_{\beta }$ maps in the
patch $U_{2}\left( z_{0},z_{1},z_{3}\right) $ and so the corresponding $%
\mathbf{v}_{2i}$ vectors. We have: 
\begin{equation}
\begin{tabular}{ll}
$r_{\alpha }=-\left\vert z_{0}\right\vert ^{2}+\left\vert z_{1}\right\vert
^{2}$ & $,$ \\ 
$r_{\beta }=2\left\vert z_{0}\right\vert ^{2}-\left\vert z_{1}\right\vert
^{2}-\left\vert z_{3}\right\vert ^{2}+t$ & ,%
\end{tabular}%
\end{equation}%
and 
\begin{equation}
\left\{ 
\begin{array}{c}
\mathbf{v}_{21}=\left( -1,2\right) \\ 
\mathbf{v}_{22}=\left( 1,-1\right) \\ 
\mathbf{v}_{23}=\left( 0,-1\right)%
\end{array}%
,\qquad \mathbf{v}_{21}+\mathbf{v}_{22}+\mathbf{v}_{23}=0\right.
\label{bsi19}
\end{equation}%
An analogous analysis for the local patch $U_{1}\left(
z_{0},z_{2},z_{3}\right) $ leads to: 
\begin{equation}
\begin{tabular}{ll}
$r_{\alpha }=2\left\vert z_{0}\right\vert ^{2}-\left\vert z_{2}\right\vert
^{2}-\left\vert z_{3}\right\vert ^{2}+t$ & , \\ 
$r_{\beta }=-\left\vert z_{0}\right\vert ^{2}+\left\vert z_{2}\right\vert
^{2}$ & ,%
\end{tabular}
\label{bsi20}
\end{equation}%
and 
\begin{equation}
\left\{ 
\begin{array}{c}
\mathbf{v}_{31}=\left( 2,-1\right) \\ 
\mathbf{v}_{32}=\left( -1,1\right) \\ 
\mathbf{v}_{33}=\left( -1,0\right)%
\end{array}%
,\qquad \mathbf{v}_{31}+\mathbf{v}_{32}+\mathbf{v}_{33}=0\right.
\label{bsi21}
\end{equation}%
The non compact direction $R$ is generated by $r_{\gamma }=\func{Im}\left(
z_{0}z_{1}z_{2}z_{3}\right) $. The special Lagrangian fiber is then
generated by the action of the three \textquotedblleft
Hamiltonians\textquotedblright\ $r_{\alpha },$ $r_{\beta }$ and $r_{\gamma }$
on $C^{3}$ via the standard symplectic form $\omega =i\sum_{i}dz_{i}\wedge d%
\overline{z}_{i}$ and the Poisson bracket $\partial z_{i}=\left\{
r,z_{i}\right\} _{\omega }$. Note that the $T^{2}$ fiber generated by $%
U\left( 1\right) \times U\left( 1\right) $ action, 
\begin{equation}
e^{i\alpha r_{\alpha }+i\beta r_{\beta }}:\left( z_{0},z_{1},z_{2}\right)
\rightarrow \left( e^{-i\left( \alpha +\beta \right) }z_{0},e^{i\alpha
}z_{1},e^{i\beta }z_{2}\right) ,  \label{bsi22}
\end{equation}%
degenerates over $z_{0}=0=z_{1}=z_{2}$.

\subsection{$NP^{3}$ as a non planar geometry}

\subsubsection{$T^{4}$\textit{\ fibration of }$NP^{3}$}

The previous analysis extends naturally to the present case. The toric
diagram of the complex projective space $P^{3}$ is given by $T^{3}$\textit{\ 
}fibration over the 3-simplex 
\begin{equation}
B_{3}:x_{1}+x_{2}+x_{3}+x_{4}=t.
\end{equation}%
The latter is a finite tetrahedron (a pyramid) embedded in $R_{+}^{4}$
parameterized by $\left( x_{1},x_{2},x_{3},x_{4}\right) $ which, in the
limit $t$ goes to zero, shrinks to the origin of $R_{+}^{4}$ where lives a $%
P^{3}$ singularity. 
\begin{figure}[tbph]
\begin{center}
\hspace{-1cm} \includegraphics[width=10cm]{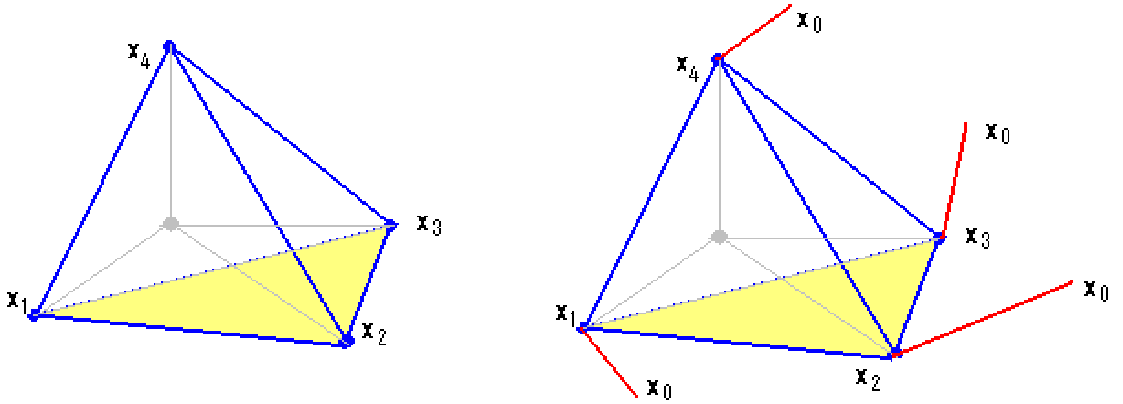}
\end{center}
\caption{{\protect\small On left: toric diagram of P}$^{3}${\protect\small .
On right toric diagram of }$\mathcal{O}\left( -4\right) \rightarrow P^{3}$.}
\end{figure}
The toric graph of the normal bundle $NP^{3}$ viewed as a $T^{4}$ fibration
over a real base $B_{4}$ is given by 
\begin{equation}
x_{0}=\frac{1}{4}\left( x_{1}+x_{2}+x_{3}+x_{4}-t\right) ,\qquad x_{i}\geq 0.
\label{bsi23}
\end{equation}%
with a $T^{4}$ fiber. Notice that $NP^{3}$ is toric; its boundary is also
toric and is given by an $T^{3}$ fibration over the real 3-spaces $D_{i}$: 
\begin{equation}
\left\{ 
\begin{array}{c}
D_{0}:x_{1}+x_{2}+x_{3}+x_{4}-t=0,\qquad x_{0}=0 \\ 
D_{1}:x_{2}+x_{3}+x_{4}-3x_{0}-t=0,\qquad x_{1}=0 \\ 
D_{2}:x_{1}+x_{3}+x_{4}-3x_{0}-t=0,\qquad x_{2}=0 \\ 
D_{3}:x_{1}+x_{2}+x_{4}-3x_{0}-t=0,\qquad x_{3}=0 \\ 
D_{4}:x_{1}+x_{2}+x_{3}-3x_{0}-t=0,\qquad x_{4}=0%
\end{array}%
\right.
\end{equation}%
The toric diagrams of $P^{3}$ and $NP^{3}$ are reported in \emph{figure }(3a)%
\emph{\ and }(3b).

\subsubsection{$T^{3}\times R$ \textit{fibration of }$NP^{3}$}

The local surface $NP^{3}$ and in general any smooth Calabi-Yau 4-fold can
be obtained by gluing together $C^{4}$ patches in a way that is consistent
with Ricci-flatness. For each $C^{4}$ patch, we have four outgoing 3-vectors
adding to zero as indicated on the example given by \emph{figure 6}, 
\begin{figure}[tbph]
\begin{center}
\hspace{-1cm} \includegraphics[width=3cm]{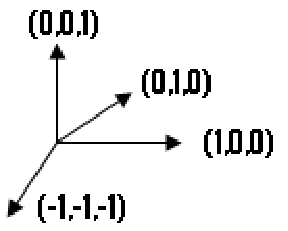}
\end{center}
\caption{The graph of $C^{4}$}
\end{figure}
\newline
The geometry of full $NP^{3}$ is then encoded in a non planar 3-dimensional
graph $\Gamma _{3}$\ in the base that corresponds to the degeneration locus
of the fibration $T^{3}\times R$. In the present example, shown on \emph{%
figure 7}, we have four (tetravalent) vertices $V_{1}$, $V_{2}$, $V_{3}$ and 
$V_{4}$. The edges $E_{i}\ $of the graph $\Gamma _{3}$ are oriented straight
lines labeled by integer 3-vectors 
\begin{equation}
\mathbf{v}_{i}=\left( p_{i},q_{i},s_{i}\right) \in Z^{3},
\end{equation}%
describing the shrinking 1-cycle $a_{i}\in H_{1}\left( T^{3}\right) $. 
\begin{figure}[tbph]
\begin{center}
\hspace{-1cm} \includegraphics[width=6cm]{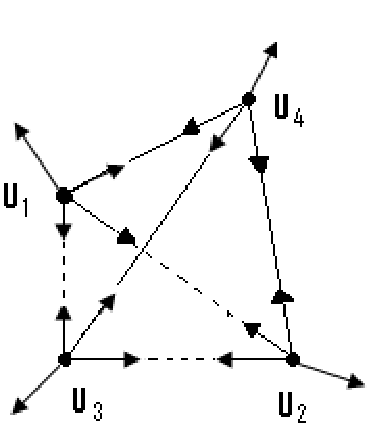}
\end{center}
\caption{{\protect\small The graph of }$T^{3}\times R${\protect\small \
fibration of }$O\left( -4\right) \rightarrow P^{3}${\protect\small . It is
built out of four }$C^{4}${\protect\small \ patches as in the figure. }}
\end{figure}
\newline
The condition of being a smooth toric Calabi-Yau is equivalent to the
condition that on each vertex $V_{i}$, if we choose the edges to be outgoing
with charges $\mathbf{v}_{i}$, we must have 
\begin{equation}
\sum_{i=1}^{4}\mathbf{v}_{i}=0.
\end{equation}%
For the case at hand, we have 
\begin{equation}
\left\{ 
\begin{array}{c}
\text{vertex }V_{1}:\text{ }\mathbf{v}_{11}+\mathbf{v}_{12}+\mathbf{v}_{13}+%
\mathbf{v}_{14}=0 \\ 
\text{vertex }V_{2}:\text{ }\mathbf{v}_{21}+\mathbf{v}_{22}+\mathbf{v}_{23}+%
\mathbf{v}_{24}=0 \\ 
\text{vertex }V_{3}:\text{ }\mathbf{v}_{31}+\mathbf{v}_{32}+\mathbf{v}_{33}+%
\mathbf{v}_{34}=0 \\ 
\text{vertex }V_{4}:\text{ }\mathbf{v}_{41}+\mathbf{v}_{42}+\mathbf{v}_{43}+%
\mathbf{v}_{44}=0%
\end{array}%
\right. .  \label{bsi25}
\end{equation}%
As we see, changing the orientation on each edge $E_{i}$ corresponds to
replacing $\mathbf{v}_{i}\rightarrow -\mathbf{v}_{i}$ which does not change
the Calabi-Yau geometry. The graph of the $T^{3}\times R$ fibration of%
\textit{\ } $NP^{3}$ involves four patches 
\begin{equation}
\begin{tabular}{llll}
$U_{1}\left( z_{0},z_{2},z_{3},z_{4}\right) \simeq $ $C^{4}$ & , & $%
U_{2}\left( z_{0},z_{1},z_{3},z_{4}\right) \simeq $ $C^{4}$ & , \\ 
$U_{3}\left( z_{0},z_{1},z_{2},z_{4}\right) \simeq $ $C^{4}$ & , & $%
U_{4}\left( z_{0},z_{1},z_{2},z_{3}\right) \simeq $ $C^{4}$ & ,%
\end{tabular}%
\end{equation}%
and can be obtained as follows:\newline
In the local patch $U_{4}\left( z_{0},z_{1},z_{2},z_{3}\right) $, the base
of the $T^{3}$ fibration is the image of the moment maps 
\begin{equation}
\left\{ 
\begin{array}{c}
r_{\alpha }=-\left\vert z_{0}\right\vert ^{2}+\left\vert z_{1}\right\vert
^{2},\qquad \\ 
r_{\beta }=-\left\vert z_{0}\right\vert ^{2}+\left\vert z_{2}\right\vert
^{2},\qquad \\ 
r_{\gamma }=-\left\vert z_{0}\right\vert ^{2}+\left\vert z_{3}\right\vert
^{2}.\qquad%
\end{array}%
\right.  \label{bsi26}
\end{equation}%
From these relations, one can read the $U^{3}\left( 1\right) $ charges $%
\left( \alpha ,\beta ,\gamma \right) $ of the $z_{i}$ variables. We have,
see also the corresponding figure, 
\begin{equation}
\left\{ 
\begin{tabular}{lll}
$\mathbf{v}_{41}$ & $=$ & $\left( -1,-1,-1\right) $ \\ 
$\mathbf{v}_{42}$ & $=$ & $\left( 1,0,0\right) $ \\ 
$\mathbf{v}_{43}$ & $=$ & $\left( 0,1,0\right) $ \\ 
$\mathbf{v}_{44}$ & $=$ & $\left( 0,0,1\right) $%
\end{tabular}%
\ ,\qquad \mathbf{v}_{41}+\mathbf{v}_{42}+\mathbf{v}_{43}+\mathbf{v}%
_{44}=0\right. .  \label{bsi27}
\end{equation}%
In the local patch $U_{1}\left( z_{0},z_{2},z_{3},z_{4}\right) $, the maps
as well as the corresponding $\mathbf{v}_{1i}$ vectors are: 
\begin{eqnarray}
&&\left\{ 
\begin{tabular}{lll}
$r_{\alpha }$ & $=$ & $3\left\vert z_{0}\right\vert ^{2}-\left\vert
z_{2}\right\vert ^{2}-\left\vert z_{3}\right\vert ^{2}-\left\vert
z_{4}\right\vert ^{2}+t$ \\ 
$r_{\beta }$ & $=$ & $-\left\vert z_{0}\right\vert ^{2}+\left\vert
z_{2}\right\vert ^{2}$ \\ 
$r_{\gamma }$ & $=$ & $-\left\vert z_{0}\right\vert ^{2}+\left\vert
z_{3}\right\vert ^{2}$%
\end{tabular}%
\ \right.  \label{bsi28} \\
&&\left\{ 
\begin{tabular}{lll}
$\mathbf{v}_{11}$ & $=$ & $\left( 3,-1,-1\right) $ \\ 
$\mathbf{v}_{12}$ & $=$ & $\left( -1,1,0\right) $ \\ 
$\mathbf{v}_{13}$ & $=$ & $\left( -1,0,1\right) $ \\ 
$\mathbf{v}_{14}$ & $=$ & $\left( -1,0,0\right) $%
\end{tabular}%
\ ,\qquad \mathbf{v}_{11}+\mathbf{v}_{12}+\mathbf{v}_{13}+\mathbf{v}%
_{14}=0\right.  \label{bsi29}
\end{eqnarray}%
Similarly, we can write down the hamiltonian maps and the vectors $\mathbf{v}%
_{2i}$ in the patch $U_{2}\left( z_{0},z_{1},z_{3},z_{4}\right) $. We have, 
\begin{eqnarray}
&&\left\{ 
\begin{tabular}{llll}
$r_{\alpha }$ & $=$ & $-\left\vert z_{0}\right\vert ^{2}+\left\vert
z_{1}\right\vert ^{2}$ & $,$ \\ 
$r_{\beta }$ & $=$ & $3\left\vert z_{0}\right\vert ^{2}-\left\vert
z_{1}\right\vert ^{2}-\left\vert z_{3}\right\vert ^{2}-\left\vert
z_{4}\right\vert ^{2}+t$ & $,$ \\ 
$r_{\gamma }$ & $=$ & $-\left\vert z_{0}\right\vert ^{2}+\left\vert
z_{3}\right\vert ^{2}$ & $.$%
\end{tabular}%
\ \right.  \label{bsi30} \\
&&\left\{ 
\begin{tabular}{lll}
$\mathbf{v}_{21}$ & $=$ & $\left( -1,3,-1\right) $ \\ 
$\mathbf{v}_{22}$ & $=$ & $\left( 1,-1,0\right) $ \\ 
$\mathbf{v}_{23}$ & $=$ & $\left( 0,-1,1\right) $ \\ 
$\mathbf{v}_{24}$ & $=$ & $\left( 0,-1,0\right) $%
\end{tabular}%
\ ,\qquad \mathbf{v}_{21}+\mathbf{v}_{22}+\mathbf{v}_{23}+\mathbf{v}%
_{24}=0\right.  \label{bsi31}
\end{eqnarray}%
We also have for the local patch $U_{3}\left( z_{0},z_{1},z_{2},z_{4}\right) 
$: 
\begin{eqnarray}
&&\left\{ 
\begin{tabular}{llll}
$r_{\alpha }$ & $=$ & $-\left\vert z_{0}\right\vert ^{2}+\left\vert
z_{1}\right\vert ^{2}$ & $,$ \\ 
$r_{\beta }$ & $=$ & $-\left\vert z_{0}\right\vert ^{2}+\left\vert
z_{2}\right\vert ^{2}$ & $,$ \\ 
$r_{\gamma }$ & $=$ & $3\left\vert z_{0}\right\vert ^{2}-\left\vert
z_{1}\right\vert ^{2}-\left\vert z_{2}\right\vert ^{2}-\left\vert
z_{4}\right\vert ^{2}+t$ & $.$%
\end{tabular}%
\ \right.  \label{bsi32} \\
&&\left\{ 
\begin{tabular}{lll}
$\mathbf{v}_{31}$ & $=$ & $\left( -1,-1,3\right) $ \\ 
$\mathbf{v}_{32}$ & $=$ & $\left( 1,0,-1\right) $ \\ 
$\mathbf{v}_{33}$ & $=$ & $\left( 0,1,-1\right) $ \\ 
$\mathbf{v}_{34}$ & $=$ & $\left( 0,0,-1\right) $%
\end{tabular}%
\ ,\qquad \mathbf{v}_{31}+\mathbf{v}_{32}+\mathbf{v}_{33}+\mathbf{v}%
_{34}=0\right. .  \label{bsi33}
\end{eqnarray}%
The non compact direction $R$ is generated by $r_{\delta }=\func{Im}\left(
z_{0}z_{1}z_{2}z_{3}z_{4}\right) $. This analysis generalizes immediately to 
$\mathcal{O}\left( -n\right) \rightarrow P^{n-1}$ with $n>4$.

\section{Local degenerate elliptic curve}

Always interested in the study of local Calabi-Yau threefold, we focus in
this section on the particular local degenerate elliptic curve, 
\begin{equation}
\begin{tabular}{llll}
$H_{3}=\mathcal{O}(+3)\oplus \mathcal{O}(-3)\rightarrow E^{\left( t,\infty
\right) }$ & $,$ & $m=3$ & $.$%
\end{tabular}
\label{m}
\end{equation}%
The elliptic curve can be generally denoted as $E^{\left( t,\mu \right) }$
as it has one Kahler parameter $t$ and one complex parameter $\mu $. Below,
we will consider the limit $\mu \rightarrow \infty $ so that $E^{\left(
t,\infty \right) }$ can be identified with $E=P_{1}^{1}\cup P_{2}^{1}\cup
P_{3}^{1}$ with matrix intersection (\ref{mi})

\subsection{Embedding local $E^{\left( t,\infty \right) }$ in NP$^{3}$}

First, notice that there exist various ways to describe the above local
elliptic curve $E^{\left( t,\infty \right) }$ in $NP^{3}$. One way to do is
to think about $NP^{3}$ as a fibration of the compact line bundle $\mathcal{O%
}\left( +3\right) $ over the local complex surface $X_{2}=\mathcal{O}\left(
-3\right) \rightarrow E^{\left( t,\infty \right) }$. In this case, the base $%
E^{\left( t,\infty \right) }$ ($E$ for short) is a toric line realized by
the special toric curve $P_{1}^{1}\cup P_{2}^{1}\cup P_{3}^{1}$. Then, $E$
is the \emph{toric boundary} of the complex projective plane $P^{2}$, 
\begin{equation}
E=\partial \left( P^{2}\right) .
\end{equation}%
So, the toric graph of the compact part of $H_{3}$ consists of three
intersecting triangles forming the boundary surface of a hollow tetrahedron.
The toric graph of the shrinking 1-cycles of $\mathcal{O}\left( +3\right)
\rightarrow X_{2}$ is given by \emph{figure 11}. 
\begin{figure}[tbph]
\begin{center}
\hspace{0cm} \includegraphics[width=4cm]{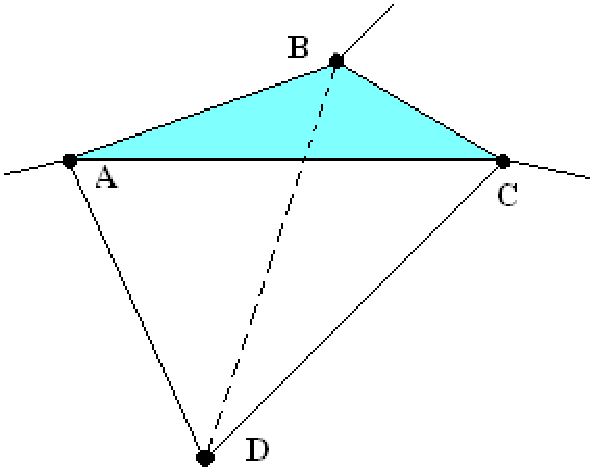}
\end{center}
\par
\vspace{-0.5 cm}
\caption{Non planar toric web-diagram of \ $\mathcal{O}\left( +3\right)
\oplus \mathcal{O}\left( -3\right) \rightarrow E^{\left( t,\infty \right) }$%
. {\protect\small This is a toric CY3 divisor of the four dimension complex
Kahler manifold} $\mathcal{O}\left( +3\right) \oplus \mathcal{O}\left(
-3\right) \rightarrow P^{2}$ \ {\protect\small The hollow triangle ABC
refers to the degenerate elliptic curve }$E^{\left( t,\infty \right) }$%
{\protect\small . The full triangles ABD, ZCD, BCD refer to the three other
projective planes. }}
\label{4v2}
\end{figure}
This non standard Calabi-Yau threefold $X_{3}=\mathcal{O}\left( +3\right)
\oplus X_{2}$, with $X_{2}=\mathcal{O}\left( -3\right) \rightarrow E$
involves non planar toric graphs. {\LARGE \newline
}Because of the nature of the base $E$ which is built out of the three
intersecting projective lines $P_{1}^{1},$ $P_{2}^{1},$ $P_{3}^{1}$ of $%
P^{2} $ and because of the $\mathcal{O}\left( +3\right) $ fibration, the
toric data of the variety $X_{3}$ can be deduced from that of the normal
bundle of the complex three dimension weighted projective space W$\mathbb{P}%
_{1113}^{3} $. \newline
An other way to thinking about $X_{3}$ is 
\begin{equation}
X_{3}=\mathcal{O}\left( -3\right) \rightarrow X_{2}^{\prime },
\end{equation}%
with 
\begin{equation}
X_{2}^{\prime }=\mathcal{O}\left( +3\right) \rightarrow E.
\end{equation}%
describing the toric boundary surface of the complex dimension three
weighted projective space; \ i.e 
\begin{equation}
X_{2}^{\prime }\subset WP_{1,1,1,3}^{3}
\end{equation}%
As we know $WP_{1,1,1,3}^{3}$, which roughly looks like $P^{3}$, has a non
planar toric graph with:\newline
- \emph{Four faces} (divisors) F$_{1}$, F$_{2}$, F$_{3}$ and F$_{4}$,\newline
- \emph{Six edges} E$_{1}$, E$_{2}$, E$_{3}$, E$_{4}$, E$_{5}$, E$_{6}$; and%
\newline
- \emph{Four vertices} V$_{1}$, V$_{2}$, V$_{3}$ and V$_{4}$.\newline
By supplying the $\mathcal{O}\left( -3\right) $ fibration, the toric
threefold $X_{3}$ can be viewed as a local Calabi-Yau \emph{submanifold} of
a complex four dimension Kahler manifold $Y_{4}$; that is: 
\begin{equation}
\begin{tabular}{lll}
$X_{3}\subset Y_{4}:$ & $Y_{4}=\mathcal{O}\left( -3\right) \rightarrow
WP_{1,1,1,3}^{3}$ & $\text{.}$%
\end{tabular}%
\end{equation}%
In the framework of the topological string setting, the 4- vertices are
described by local patches $U_{i}\simeq $ $C^{4}$, \newline
\begin{figure}[tbph]
\begin{center}
\hspace{0cm} \includegraphics[width=10cm]{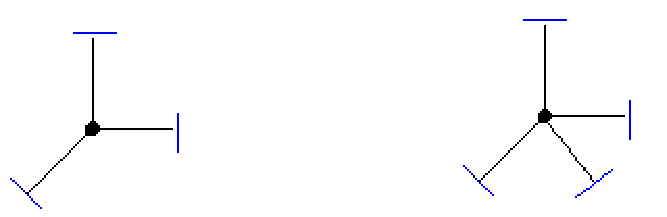}
\end{center}
\caption{{\protect\small On left the graph of topological 3-vertex where
each leg ends on a lagrangian submanifold. On right, we have the 4-vertex
analog.}}
\end{figure}
\ \newline
We will address this question with some details later on; but before that we
first focus on the toric data of 
\begin{equation}
\mathcal{O}\left( -3\right) \rightarrow WP_{1,1,1,3}^{3}
\end{equation}%
and its toric submanifold 
\begin{equation}
\mathcal{O}\left( -3\right) \rightarrow \partial \left(
WP_{1,1,1,3}^{3}\right) .
\end{equation}%
Generally, a toric CY4- fold with local patches $C^{4}$ parameterized by
complex coordinates $\left\{ w_{i}\right\} $ has the natural toric (trivial)
fibration 
\begin{equation}
B_{4}\times T^{4},
\end{equation}%
with real base $B_{4}\subset R^{4}$ and a Kahler form $J=i\sum_{i,j=1}^{4}\
dw_{i}\wedge d\overline{w}_{i}$. This form splits in the $\left( \left\vert
w_{i}\right\vert ^{2},\vartheta _{i}\right) $ polar coordinates as 
\begin{equation}
\begin{tabular}{llllll}
$\mathbf{J}=\sum_{i=1}^{4}d\rho _{i}^{2}\wedge d\vartheta _{i}$ & $,$ & $%
\rho _{i}^{2}=\left\vert w_{i}\right\vert ^{2}$ & $,$ & $w_{i}=\rho
_{i}e^{i\vartheta _{i}}$ & $.$%
\end{tabular}
\label{an}
\end{equation}%
A Lagrangian submanifold $L_{4}$ of $C^{4}$ is a real 4-dimensional subspace
satisfying the usual property, 
\begin{equation}
\mathbf{J}\mid _{L_{4}}=0.  \label{ad}
\end{equation}%
By using eq(\ref{an}), we see that this constraint eq can be solved in
different ways; for instance by taking $\rho _{i}^{2}$ = constant ($d\rho
_{i}^{2}=0$) or by setting $\vartheta _{i}$ = constant ($d\vartheta _{i}=0$%
). \newline
One can also build special Lagrangian submanifolds $\mathcal{L}$ satisfying,
in addition to eq(\ref{ad}), the constraint eq 
\begin{equation}
\left( \func{Im}\Omega \right) \mid _{\mathcal{L}}=0
\end{equation}%
where $\Omega $ is the usual holomorphic $\left( 4,0\right) $ form. \newline
In the present study, we are interested in the $T^{2}\times R$ special
Lagrangian fibration of the toric Calabi-Yau threefold $\mathcal{O}\left(
-3\right) \rightarrow \partial \left( WP_{1,1,1,3}^{3}\right) $. This
fibration extends to a $T^{3}\times R$ fibration of the ambient space $%
\mathcal{O}\left( -3\right) \rightarrow WP_{1,1,1,3}^{3}$.

\subsubsection{Toric graph of $\mathcal{O}\left( -3\right) \rightarrow
\partial \left( WP_{1,1,1,3}^{3}\right) $}

The toric graph of the Calabi-Yau threefold $\mathcal{O}\left( -3\right)
\rightarrow \partial \left( WP_{1,1,1,3}^{3}\right) $ can be determined from
the graph of the manifold $\mathcal{O}\left( -3\right) \rightarrow
WP_{1,1,1,3}^{3}$. Both of these Kahler manifolds are realized as a complex
3- and \ complex 4- dimension toric hypersurfaces embedded in $C^{5}$. For
the case of $\mathcal{O}\left( -3\right) \rightarrow WP_{1,1,1,3}^{3}$, the
defining equation is given by 
\begin{equation}
\sum_{i=1}^{3}|w_{i}|^{2}+3|w_{4}|^{2}-3|w_{0}|^{2}=t.  \label{h}
\end{equation}%
For $O\left( -3\right) \rightarrow \partial \left( WP_{1,1,1,3}^{3}\right) $
we have, in addition to \ref{h}, the extra condition 
\begin{equation}
w_{1}w_{2}w_{3}=0.  \label{hh}
\end{equation}%
As mentioned earlier, eq(\ref{hh}) may be solved in three ways; either by $%
w_{1}=0$ whatever $w_{2}$ and $w_{3}$ are, or by taking$\ w_{2}=0$ or again
by setting $w_{3}=0$. Notice that both eqs(\ref{h}-\ref{hh}) are invariant
under the $U\left( 1\right) $ transformations of the complex variables 
\begin{equation}
w_{j}\equiv e^{i\alpha q_{j}}w_{j}.
\end{equation}%
Notice also that the sum of the $U\left( 1\right) $ charges 
\begin{equation}
\left( q_{0},q_{1},q_{2},q_{3},q_{4}\right) =\left( -3,1,1,1,3\right)
\end{equation}%
is non zero; 
\begin{equation}
\sum_{j=0}^{4}q_{j}=3.  \label{cyc}
\end{equation}%
It shows that $O\left( -3\right) \rightarrow WP_{1,1,1,3}^{3}$ is not a
Calabi-Yau 4-fold; while $\mathcal{O}\left( -3\right) \rightarrow \partial
\left( WP_{1,1,1,3}^{3}\right) $ is a Calabi-Yau threefold. To handle the
relations (\ref{h}-\ref{hh}), we shall proceed as follows: First deal with
eq(\ref{h}) and then implement the constraint eq(\ref{hh}).

\subsubsection{Analysis of eq(\protect\ref{h})}

We start from equation $%
\sum_{i=1}^{3}|w_{i}|^{2}+3|w_{4}|^{2}-3|w_{0}|^{2}=t $ and solve it in four
different ways according to which set of variables is used. We have the
following $C^{4}$ patches: 
\begin{equation}
\begin{tabular}{ll}
$U_{1}=U_{1}\left( w_{2},w_{3},w_{4},w_{0}\right) \ $ & , \\ 
$U_{2}=U_{2}\left( w_{1},w_{3},w_{4},w_{0}\right) \ $ & , \\ 
$U_{3}=U_{3}\left( w_{1},w_{2},w_{4},w_{0}\right) \ $ & , \\ 
$U_{4}=U_{4}\left( w_{1},w_{2},w_{3},w_{0}\right) \ $ & .%
\end{tabular}
\label{c4}
\end{equation}%
On the coordinate patch $U_{1}$, the Kahler manifold $\mathcal{O}\left(
-3\right) \rightarrow WP_{1,1,1,3}^{3}$ is described by the codimension one
hypersurface of $C^{5}$, 
\begin{equation}
|w_{4}|^{2}=\frac{t}{3}+|w_{0}|^{2}-\frac{1}{3}\sum_{i=1}^{3}|w_{i}|^{2}.
\end{equation}%
Similarly, we have for the patch $U_{1}$, $U_{2}$ and $U_{3}$ the following
relations, 
\begin{eqnarray}
|w_{1}|^{2} &=&t+3|w_{0}|^{2}-|w_{2}|^{2}-|w_{3}|^{2}-3|w_{4}|^{2},  \notag
\\
|w_{2}|^{2} &=&t+3|w_{0}|^{2}-|w_{1}|^{2}-|w_{3}|^{2}-3|w_{4}|^{2}, \\
|w_{3}|^{2} &=&t+3|w_{0}|^{2}-|w_{1}|^{2}-|w_{2}|^{2}-3|w_{4}|^{2}.  \notag
\end{eqnarray}%
Each one of the local patches $U_{i}$ is isomorphic to \QTR{cal}{C}$^{4}$.
To get the toric graph of the shrinking 2- cycles, we have to first identify
the hamiltonians of the $T^{3}\times R$ fibration of $\mathcal{O}\left(
-3\right) \rightarrow WP_{1,1,1,3}^{3}$.

\subsubsection{Hamiltonians on the U$_{4}$ patch}

Following the same method, we have used in subsection 2.2 concerning $%
T^{3}\times R$ fibration of the 4- fold $\mathcal{O}\left( -4\right)
\rightarrow P^{3}$, \textrm{see eqs(\ref{h}-\ref{hh})}, the hamiltonians 
\begin{equation}
H_{\alpha },\quad H_{\beta },\quad H_{\gamma },\quad H_{\delta },
\end{equation}%
generating of the $T^{3}\times R$ special Lagrangian fibration depends on
the local patches we are sitting on. We have:

\textbf{Local patch} $w_{4}=f_{4}\left( w_{i}\right) $: \ \newline
On the patch $U_{4}\left( w_{1},w_{2},w_{3},w_{0}\right) $ of the 4- fold $%
\mathcal{O}\left( -3\right) \rightarrow WP_{1,1,1,3}^{3}$ where $w_{4}$ is
solved in terms of the four complex variables $\left(
w_{1},w_{2},w_{3},w_{0}\right) $ as shown above, the three projective
variables $w_{1},w_{2},w_{3}$ play a symmetric role. So the three
hamiltonians 
\begin{equation}
H_{\alpha }=H_{\alpha }^{\left( 4\right) },\quad H_{\beta }=H_{\beta
}^{\left( 4\right) },\quad H_{\gamma }=H_{\gamma }^{\left( 4\right) },
\end{equation}%
generating the 2-cycles of $T^{3}$ can be written as follows: 
\begin{equation}
U_{4}:\left\{ 
\begin{array}{c}
\begin{array}{c}
H_{\alpha }^{\left( 4\right) }=|w_{1}|^{2}-|w_{0}|^{2} \\ 
H_{\beta }^{\left( 4\right) }=|w_{2}|^{2}-|w_{0}|^{2} \\ 
H_{\gamma }^{\left( 4\right) }=|w_{3}|^{2}-|w_{0}|^{2}%
\end{array}
\\ 
\mathrm{H}_{\delta }^{\left( 4\right) }=\func{Im}w_{1}w_{2}w_{3}w_{4}w_{0}%
\end{array}%
\right. .
\end{equation}%
From these relations, we can write down the outgoing ("momentum") vectors 
\begin{equation}
v_{1}^{\left( 4\right) },\quad v_{2}^{\left( 4\right) },\quad v_{3}^{\left(
4\right) },\quad v_{0}^{\left( 4\right) },
\end{equation}%
describing the shrinking 1-cycles on the four edges of the vertex $C^{4}$
associated with the toric graph of the $U_{4}$ patch. We have, 
\begin{equation*}
\begin{tabular}{lllll}
$v_{1}^{\left( 4\right) }=\left( 1,0,0\right) ,$ & $v_{2}^{\left( 4\right)
}=\left( 0,1,0\right) ,$ & $v_{3}^{\left( 4\right) }=\left( 0,0,1\right) ,$
& $v_{0}^{\left( 4\right) }=\left( -1,-1,-1\right) $ & .%
\end{tabular}%
\end{equation*}%
Before proceeding further, notice the two following: \newline
(\textbf{i}) the sum of the $v_{i}^{\left( 4\right) }$ momenta add exactly
to zero 
\begin{equation}
\sum_{i=0}^{3}v_{i}^{\left( 4\right) }=\left( 0,0,0\right)
\end{equation}%
This property seems a little bit strange as one expects something different
from zero since the 4- fold $\mathcal{O}\left( -3\right) \rightarrow
WP_{1,1,1,3}^{3}$ is not a Calabi-Yau manifold. We will turn later on to
this point and show that for the other vertices the sum of momentum vectors
is non zero. \newline
(\textbf{ii}) Imposing the condition 
\begin{equation}
w_{1}w_{2}w_{3}=0,
\end{equation}%
on the patch $U_{4}$ and solving it in three ways as $w_{1}=0$, $%
w_{2}w_{3}\neq 0$, or $w_{2}=0$, $w_{1}w_{3}\neq 0$ or again $w_{3}=0$, $%
w_{2}w_{1}\neq 0$, we discover that $U_{4}$ can be split into complex 3-
dimension local patches $U_{4i}$ equivalent to $C^{3}$. These are given by, 
\begin{eqnarray}
U_{41} &=&U_{41}\left( w_{2},w_{3},w_{0}\right) ,  \notag \\
U_{42} &=&U_{42}\left( w_{1},w_{3},w_{0}\right) , \\
U_{43} &=&U_{43}\left( w_{1},w_{2},w_{0}\right) ,  \notag
\end{eqnarray}%
where, for instance, the subindex on $U_{41}$ refers to the fact that on
this local patch we have $w_{4}=f\left( w_{i}\right) $ and $w_{1}=0$. The
hamiltonians generating the 1- cycles of the $T^{2}\times R$ fibration read
respectively as follows: 
\begin{eqnarray}
U_{41}\ &:&\left\{ 
\begin{array}{c}
H_{\beta }^{\left( 41\right) }=|w_{2}|^{2}-|w_{0}|^{2} \\ 
H_{\gamma }^{\left( 41\right) }=|w_{3}|^{2}-|w_{0}|^{2}%
\end{array}%
\right. ,\quad  \notag \\
U_{42} &:&\left\{ 
\begin{array}{c}
H_{\alpha }^{\left( 42\right) }=|w_{1}|^{2}-|w_{0}|^{2} \\ 
H_{\gamma }^{\left( 42\right) }=|w_{3}|^{2}-|w_{0}|^{2}%
\end{array}%
\right. ,\quad \\
U_{43} &:&\left\{ 
\begin{array}{c}
H_{\alpha }^{\left( 43\right) }=|w_{1}|^{2}-|w_{0}|^{2} \\ 
H_{\beta }^{\left( 43\right) }=|w_{2}|^{2}-|w_{0}|^{2}%
\end{array}%
\right.  \notag
\end{eqnarray}%
From these relations, we can write down the momentum vectors of the
shrinking 1- cycles of the $T^{2}\times R$ fibration. We have the following
projections: 
\begin{equation}
\begin{tabular}{llll}
$U_{41}:$ & $v_{2}^{\left( 41\right) }=\left( \mathrm{\ast },1,0\right) ,$ & 
$v_{3}^{\left( 41\right) }=\left( \mathrm{\ast },0,1\right) ,$ & $%
v_{0}^{\left( 41\right) }=\left( \mathrm{\ast },-1,-1\right) ,$ \\ 
$U_{42}:$ & $v_{1}^{\left( 42\right) }=\left( 1,\mathrm{\ast },0\right) ,$ & 
$v_{3}^{\left( 42\right) }=\left( 0,\mathrm{\ast },1\right) ,$ & $%
v_{0}^{\left( 42\right) }=\left( -1,\mathrm{\ast },-1\right) ,$ \\ 
$U_{43}:$ & $v_{1}^{\left( 43\right) }=\left( 1,0,\mathrm{\ast }\right) ,$ & 
$v_{2}^{\left( 43\right) }=\left( 0,1,\mathrm{\ast }\right) ,$ & $%
v_{0}^{\left( 43\right) }=\left( -1,-1,\mathrm{\ast }\right) .$%
\end{tabular}%
\end{equation}%
Note that on all $U_{4i}$ patches, we have 
\begin{equation}
\mathrm{H}_{\delta }^{\left( 4i\right) }=0
\end{equation}%
and the sums $\sum_{j}v_{j}^{\left( 4i\right) }$ add exactly to zero.

\subsection{More on Hamiltonians}

\textbf{Local patch\ }$U_{1}$:\qquad \newline
On the local patch $U_{1}\left( w_{2},w_{3},w_{4},w_{0}\right) $, the
variable $w_{1}$ is expressed as $f_{1}\left( w_{2},w_{3},w_{4},w_{0}\right) 
$. The method is quite similar to the one used above. The hamiltonians
generating the cycles of the $T^{3}\times R$ of the $U_{1}$ patch of the
complex four dimension space $\mathcal{O}\left( -3\right) \rightarrow
WP_{1,1,1,3}^{3}$ read as follows: 
\begin{equation}
U_{1}\ :\left\{ 
\begin{array}{c}
H_{\alpha }^{\left( 1\right)
}=-|w_{2}|^{2}-|w_{3}|^{2}-3|w_{4}|^{2}+2|w_{0}|^{2}+t \\ 
H_{\beta }^{\left( 1\right) }=|w_{2}|^{2}-|w_{0}|^{2}\text{ \ \ \ \ \ \ \ \
\ \ \ \ \ \ \ \ \ \ \ \ \ \ \ \ \ \ \ \ \ \ \ \ \ } \\ 
H_{\gamma }^{\left( 1\right) }=|w_{3}|^{2}-|w_{0}|^{2}\text{ \ \ \ \ \ \ \ \
\ \ \ \ \ \ \ \ \ \ \ \ \ \ \ \ \ \ \ \ \ \ \ \ \ \ }%
\end{array}%
\right. ,
\end{equation}%
where we have substituted $w_{1}$ by its expression $f_{1}\left(
w_{2},w_{3},w_{4},w_{0}\right) $. The momentum vectors 
\begin{equation}
v_{i}^{\left( 1\right) },\quad i=0,1,2,3,
\end{equation}%
associated with the shrinking \textrm{2- cycles} of the special Lagrangian
fibration are given by: 
\begin{equation}
\begin{tabular}{llll}
$v_{1}^{\left( 1\right) }=\left( -1,1,0\right) $ & $,$ & $v_{2}^{\left(
1\right) }=\left( -1,0,1\right) $ & $,$ \\ 
$v_{3}^{\left( 1\right) }=\left( -3,0,0\right) $ & $,$ & $v_{0}^{\left(
1\right) }=\left( 2,-1,-1\right) $ & $.$%
\end{tabular}%
\end{equation}%
Note that, contrary to the previous case, the sum of these vectors is non
zero. 
\begin{equation}
\sum_{i=0}^{3}v_{i}^{\left( 1\right) }=\left( -3,0,0\right) .
\end{equation}%
This property was expected and it reflects just the fact that complex
4-dimension $\mathcal{O}\left( -3\right) \rightarrow WP_{1,1,1,3}^{3}$
described by the hypersurface, 
\begin{equation}
\sum_{i=1}^{3}|w_{i}|^{2}+3|w_{4}|^{2}-3|w_{0}|^{2}=t,
\end{equation}%
is not a Calabi-Yau 4-fold. Moreover solving the\textrm{\ }condition\textrm{%
\ }$w_{1}w_{2}w_{3}=0$\textrm{\ }on the patch\textrm{\ }$U_{1}$\textrm{\ }as 
\begin{equation}
\begin{tabular}{llll}
$w_{2}=0$ & or & $w_{3}=0$ & ,%
\end{tabular}%
\end{equation}%
\textrm{\ } \textrm{\ }we find that\textrm{\ }$U_{1}$\textrm{\ }can be split
into two local\textrm{\ }patches\textrm{\ }$U_{1i}$\textrm{\ }equivalent to%
\textrm{\ }$C^{3}$.\textrm{\ }These are given by, 
\begin{equation}
\begin{tabular}{ll}
$U_{12}=U_{12}\left( w_{3},w_{4},w_{0}\right) $ & , \\ 
$U_{13}=U_{13}\left( w_{2},w_{4},w_{0}\right) $ & .%
\end{tabular}%
\end{equation}%
Then, the corresponding hamiltonians $H_{\alpha }^{\left( 12\right)
},H_{\gamma }^{\left( 12\right) }$ for the case $w_{2}=0$ and $H_{\alpha
}^{\left( 13\right) },$ $H_{\beta }^{\left( 13\right) }$ for $w_{3}=0$, read
as follows: 
\begin{eqnarray}
U_{12}\ &:&\left\{ 
\begin{array}{c}
H_{\alpha }^{\left( 12\right) }=-|w_{3}|^{2}-3|w_{4}|^{2}+2|w_{0}|^{2}+t \\ 
H_{\gamma }^{\left( 12\right) }=|w_{3}|^{2}-|w_{0}|^{2}\text{\qquad \qquad
\qquad\ \ \ }%
\end{array}%
\right.  \notag \\
U_{13}\ &:&\left\{ 
\begin{array}{c}
H_{\alpha }^{\left( 13\right) }=-|w_{2}|^{2}-3|w_{4}|^{2}+2|w_{0}|^{2}+t \\ 
H_{\beta }^{\left( 13\right) }=|w_{2}|^{2}-|w_{0}|^{2}\text{ \ \ \ \ \ \ \ \
\ \ \ \ \ \ \ \ \ \ \ \ \ }%
\end{array}%
\right. .
\end{eqnarray}%
and they generate 1-cycles of the $T^{2}\times R$ ,. The vector momenta of
the shrinking cycles are 
\begin{eqnarray}
\ v_{3}^{\left( 12\right) } &=&\left( -1,\mathrm{\ast },1\right) ,\quad
v_{4}^{\left( 12\right) }=\left( -3,\mathrm{\ast },0\right) ,\quad
v_{0}^{\left( 12\right) }=\left( 2,\mathrm{\ast },-1\right) ,  \notag \\
v_{2}^{\left( 13\right) } &=&\left( -1,1,\mathrm{\ast }\right) ,\quad
v_{4}^{\left( 13\right) }=\left( -3,0,\mathrm{\ast }\right) ,\quad
v_{0}^{\left( 13\right) }=\left( 2,-1,\mathrm{\ast }\right) .
\end{eqnarray}

\textbf{Local patch }$U_{2}$:\newline
Similarly, the hamiltonians generating the cycles of the $T^{3}\times R$
fibration on the $U_{2}=U_{2}\left( w_{1},w_{3},w_{4},w_{0}\right) $ patch,
with $w_{2}=f_{2}\left( w_{i}\right) $, read as follows: 
\begin{equation}
U_{2}:\left\{ 
\begin{tabular}{lll}
$H_{\alpha }^{\left( 1\right) }$ & $=$ & $|w_{1}|^{2}-|w_{0}|^{2}$ \\ 
$H_{\beta }^{\left( 1\right) }$ & $=$ & $%
t+2|w_{0}|^{2}-|w_{1}|^{2}-|w_{3}|^{2}-3|w_{4}|^{2}$ \\ 
$H_{\gamma }^{\left( 1\right) }$ & $=$ & $|w_{3}|^{2}-|w_{0}|^{2}$%
\end{tabular}%
\ \right. .
\end{equation}%
The vectors 
\begin{equation}
v_{0}^{\left( 2\right) },\text{ \ }v_{1}^{\left( 2\right) },\text{ \ }%
v_{2}^{\left( 2\right) },\text{ \ }v_{3}^{\left( 2\right) },\text{ }
\end{equation}%
associated with the shrinking 1-cycles of these hamiltonians are 
\begin{eqnarray}
v_{1}^{\left( 2\right) } &=&\left( 1,-1,0\right) ,\quad v_{2}^{\left(
2\right) }=\left( 0,-1,1\right) ,  \notag \\
v_{3}^{\left( 2\right) } &=&\left( 0,-3,0\right) ,\quad v_{0}^{\left(
2\right) }=\left( -1,2,-1\right) .
\end{eqnarray}%
Here also, the sum of the $v_{i}^{\left( 2\right) }$s is non zero 
\begin{equation}
\sum_{i=0}^{3}v_{i}^{\left( 2\right) }=\left( 0,-3,0\right) .
\end{equation}

\textbf{Local patch }$U_{3}$:\newline
In this case, the Hamiltonians read as, 
\begin{equation}
U_{3}\ :\left\{ 
\begin{tabular}{lll}
$H_{\alpha }^{\left( 1\right) }$ & $=$ & $|w_{1}|^{2}-|w_{0}|^{2}$ \\ 
$H_{\beta }^{\left( 1\right) }$ & $=$ & $|w_{2}|^{2}-|w_{0}|^{2}$ \\ 
$H_{\gamma }^{\left( 1\right) }$ & $=$ & $%
t+2|w_{0}|^{2}-|w_{1}|^{2}-|w_{2}|^{2}-3|w_{4}|^{2}$%
\end{tabular}
\right.
\end{equation}
and the associated vectors $v_{i}^{\left( 3\right) }$ take the form: 
\begin{eqnarray}
v_{1}^{\left( 3\right) } &=&\left( 1,0,-1\right) ,\quad v_{2}^{\left(
3\right) }=\left( 0,1,-1\right) ,\quad  \notag \\
v_{3}^{\left( 3\right) } &=&\left( 0,0,-3\right) ,\quad v_{0}^{\left(
3\right) }=\left( -1,-1,2\right) .
\end{eqnarray}
The sum over the $v_{i}^{\left( 3\right) }$'s is non zero and reads as 
\begin{equation}
\sum_{i=0}^{3}v_{i}^{\left( 3\right) }=\left( 0,0,-3\right) .
\end{equation}

\textbf{Implementing the constraint} eq(\ref{hh})\newline
A way to get the Hamiltonians of the fibration and the toric data for $%
\mathcal{O}\left( -3\right) \rightarrow \partial \left(
WP_{1,1,1,3}^{3}\right) $ is to start from the Hamiltonians of 
\begin{equation}
\mathcal{O}\left( -3\right) \rightarrow WP_{1,1,1,3}^{3},
\end{equation}%
and implement the constraint eq 
\begin{equation}
w_{1}w_{2}w_{3}=0.  \label{css}
\end{equation}%
But to make direct contact with the toric analysis of \textrm{\cite{vafa}}
for the topological 3-vertex of $\mathcal{O}\left( -3\right) \rightarrow
P^{2}$, it is interesting to consider separately the solutions $w_{1}=0$,$\
w_{2}\ =0$ and $w_{3}=0$ of the constraint eq(\ref{css}).

\emph{Divisor} $w_{1}=0:\qquad \ $\newline
Setting $w_{1}=0$ in eq(\ref{h}), we get the complex 3- dimension divisor 
\begin{equation}
D_{1}:\ |w_{2}|^{2}+|w_{3}|^{2}+3|w_{4}|^{2}-3|w_{0}|^{2}=t.
\end{equation}%
This complex Kahler 3-fold can be covered by three patches 
\begin{eqnarray}
D_{1}^{\left( 4\right) } &=&D_{1}^{\left( 4\right) }\left(
w_{2},w_{3},w_{0}\right) ,  \notag \\
D_{1}^{\left( 3\right) } &=&D_{1}^{\left( 3\right) }\left(
w_{2},w_{4},w_{0}\right) , \\
D_{1}^{\left( 2\right) } &=&D_{1}^{\left( 2\right) }\left(
w_{3},w_{4},w_{0}\right) .  \notag
\end{eqnarray}%
The hamiltonians of the $T_{1}^{2}\times R$ fibration of these patches, with 
\begin{equation}
T_{1}^{2}=S_{\beta }^{1}\times S_{\gamma }^{1},
\end{equation}%
and $\alpha $\ and $\beta $\ referring to the group parameters, read as
follows: 
\begin{eqnarray}
D_{1}^{\left( 4\right) } &:&\left\{ 
\begin{tabular}{lll}
$h_{1\beta }^{\left( 4\right) }$ & $=$ & $\left\vert w_{2}\right\vert
^{2}-\left\vert w_{0}\right\vert ^{2}$ \\ 
$h_{1\gamma }^{\left( 4\right) }$ & $=$ & $\left\vert w_{3}\right\vert
^{2}-\left\vert w_{0}\right\vert ^{2}$%
\end{tabular}%
\ \right.  \notag \\
D_{1}^{\left( 3\right) } &:&\left\{ 
\begin{tabular}{lll}
$h_{1\beta }^{\left( 4\right) }$ & $=$ & $\left\vert w_{2}\right\vert
^{2}-\left\vert w_{0}\right\vert ^{2}$ \\ 
$h_{1\gamma }^{\left( 4\right) }$ & $=$ & $2\left\vert w_{0}\right\vert
^{2}-3|w_{4}|^{2}-|w_{2}|^{2}+t$%
\end{tabular}%
\ \right. \\
D_{1}^{\left( 2\right) } &:&\left\{ 
\begin{tabular}{lll}
$h_{1\beta }^{\left( 4\right) }$ & $=$ & $2\left\vert w_{0}\right\vert
^{2}-3|w_{4}|^{2}-|w_{3}|^{2}+t$ \\ 
$h_{1\gamma }^{\left( 4\right) }$ & $=$ & $\left\vert w_{3}\right\vert
^{2}-\left\vert w_{0}\right\vert ^{2}$%
\end{tabular}%
\ \right. .  \notag
\end{eqnarray}%
The momentum vectors of the shrinking 1-cycles\ read then as: 
\begin{equation}
\begin{tabular}{llllll}
$v_{12}^{\left( 4\right) }=\left( \mathrm{\ast },1,0\right) $ & $,$ & $%
v_{13}^{\left( 4\right) }=\left( \mathrm{\ast },0,1\right) $ & $,$ & $%
v_{10}^{\left( 4\right) }=\left( \mathrm{\ast },-1,-1\right) $ & , \\ 
$v_{12}^{\left( 3\right) }=\left( \mathrm{\ast },1,-1\right) $ & , & $%
v_{14}^{\left( 3\right) }=\left( \mathrm{\ast },0,-3\right) $ & , & $%
v_{10}^{\left( 3\right) }=\left( \mathrm{\ast },-1,2\right) $ & , \\ 
$v_{13}^{\left( 2\right) }=\left( \mathrm{\ast },-1,1\right) $ & , & $%
v_{14}^{\left( 2\right) }=\left( \mathrm{\ast },-3,0\right) $ & , & $%
v_{10}^{\left( 2\right) }=\left( \mathrm{\ast },2,-1\right) $ & .%
\end{tabular}%
\end{equation}

\emph{Divisor} $w_{2}=0:\qquad \ $\newline
Setting $w_{2}=0$ in eq(\ref{h}), we get the complex 3- dimension divisor 
\begin{equation}
D_{2}:|w_{1}|^{2}\ +|w_{3}|^{2}+3|w_{4}|^{2}-3|w_{0}|^{2}=t.
\end{equation}%
This complex Kahler 3-fold can be covered by three patches 
\begin{eqnarray}
D_{2}^{\left( 4\right) } &=&D_{2}^{\left( 4\right) }\left(
w_{1},w_{3},w_{0}\right) ,  \notag \\
D_{2}^{\left( 3\right) } &=&D_{2}^{\left( 3\right) }\left(
w_{1},w_{4},w_{0}\right) , \\
D_{2}^{\left( 1\right) } &=&D_{2}^{\left( 1\right) }\left(
w_{3},w_{4},w_{0}\right) .  \notag
\end{eqnarray}%
The hamiltonians of the $T_{2}^{2}\times R$ fibration, with $%
T_{2}^{2}=S_{\alpha }^{1}\times S_{\gamma }^{1}$, of these patches read as
follows: 
\begin{eqnarray}
D_{2}^{\left( 4\right) } &:&\left\{ 
\begin{tabular}{lll}
$h_{1\alpha }^{\left( 4\right) }$ & $=$ & $\left\vert w_{1}\right\vert
^{2}-\left\vert w_{0}\right\vert ^{2}$ \\ 
$h_{1\gamma }^{\left( 4\right) }$ & $=$ & $\left\vert w_{3}\right\vert
^{2}-\left\vert w_{0}\right\vert ^{2}$%
\end{tabular}%
\ \right.  \notag \\
D_{2}^{\left( 3\right) } &:&\left\{ 
\begin{tabular}{lll}
$h_{1\alpha }^{\left( 4\right) }$ & $=$ & $\left\vert w_{1}\right\vert
^{2}-\left\vert w_{0}\right\vert ^{2}$ \\ 
$h_{1\gamma }^{\left( 4\right) }$ & $=$ & $2\left\vert w_{0}\right\vert
^{2}-3|w_{4}|^{2}-|w_{1}|^{2}+t$%
\end{tabular}%
\ \right. \\
D_{2}^{\left( 1\right) } &:&\left\{ 
\begin{tabular}{lll}
$h_{1\alpha }^{\left( 4\right) }$ & $=$ & $2\left\vert w_{0}\right\vert
^{2}-3|w_{4}|^{2}-|w_{3}|^{2}+t$ \\ 
$h_{1\gamma }^{\left( 4\right) }$ & $=$ & $\left\vert w_{3}\right\vert
^{2}-\left\vert w_{0}\right\vert ^{2}$%
\end{tabular}%
\ \right. .  \notag
\end{eqnarray}%
The momentum vectors of the shrinking 1-cycles\ read then as: 
\begin{equation*}
\begin{tabular}{llllll}
$v_{21}^{\left( 4\right) }=\left( 1,\mathrm{\ast },0\right) $ & , & $%
v_{23}^{\left( 4\right) }=\left( 0,\mathrm{\ast },1\right) $ & , & $%
v_{20}^{\left( 4\right) }=\left( -1,\mathrm{\ast },-1\right) $ & , \\ 
$v_{21}^{\left( 3\right) }=\left( 1,\mathrm{\ast },-1\right) $ & , & $%
v_{24}^{\left( 3\right) }=\left( 0,\mathrm{\ast },-3\right) $ & , & $%
v_{20}^{\left( 3\right) }=\left( -1,\mathrm{\ast },2\right) $ & , \\ 
$v_{23}^{\left( 1\right) }=\left( -1,\mathrm{\ast },1\right) $ & , & $%
v_{24}^{\left( 1\right) }=\left( -3,\mathrm{\ast },0\right) $ & , & $%
v_{20}^{\left( 1\right) }=\left( 2,\mathrm{\ast },-1\right) $ & .%
\end{tabular}%
\end{equation*}

\emph{Divisor} $w_{3}=0:\qquad \ $\newline
Setting $w_{3}=0$ in eq(\ref{h}), we get the complex 3- dimension divisor 
\begin{equation}
D_{3}:|w_{1}|^{2}+|w_{2}|^{2}\ +3|w_{4}|^{2}-3|w_{0}|^{2}=t
\end{equation}%
This complex Kahler 3-fold can be covered by three patches 
\begin{eqnarray}
D_{3}^{\left( 4\right) } &=&D_{3}^{\left( 4\right) }\left(
w_{1},w_{2},w_{0}\right) ,  \notag \\
D_{3}^{\left( 2\right) } &=&D_{3}^{\left( 2\right) }\left(
w_{1},w_{4},w_{0}\right) , \\
D_{3}^{\left( 1\right) } &=&D_{3}^{\left( 1\right) }\left(
w_{2},w_{4},w_{0}\right) .  \notag
\end{eqnarray}%
The hamiltonians of the $T_{3}^{2}\times R$ fibration, with $%
T_{3}^{2}=S_{\alpha }^{1}\times S_{\beta }^{1}$, of these patches read as
follows: 
\begin{eqnarray}
D_{3}^{\left( 4\right) } &:&\left\{ 
\begin{tabular}{lll}
$h_{3\alpha }^{\left( 4\right) }$ & $=$ & $\left\vert w_{1}\right\vert
^{2}-\left\vert w_{0}\right\vert ^{2}$ \\ 
$h_{3\beta }^{\left( 4\right) }$ & $=$ & $\left\vert w_{2}\right\vert
^{2}-\left\vert w_{0}\right\vert ^{2}$%
\end{tabular}%
\ \right. ,  \notag \\
D_{3}^{\left( 2\right) } &:&\left\{ 
\begin{tabular}{lll}
$h_{3\alpha }^{\left( 2\right) }$ & $=$ & $\left\vert w_{1}\right\vert
^{2}-\left\vert w_{0}\right\vert ^{2}$ \\ 
$h_{3\beta }^{\left( 2\right) }$ & $=$ & $2\left\vert w_{0}\right\vert
^{2}-3|w_{4}|^{2}-|w_{1}|^{2}+t$%
\end{tabular}%
\ \right. , \\
D_{3}^{\left( 1\right) } &:&\left\{ 
\begin{tabular}{lll}
$h_{3\alpha }^{\left( 1\right) }$ & $=$ & $2\left\vert w_{0}\right\vert
^{2}-3|w_{4}|^{2}-|w_{2}|^{2}+t$ \\ 
$h_{3\beta }^{\left( 1\right) }$ & $=$ & $\left\vert w_{2}\right\vert
^{2}-\left\vert w_{0}\right\vert ^{2}$%
\end{tabular}%
\ \right. .  \notag
\end{eqnarray}%
The momentum vectors of the shrinking 1-cycles\ read then as: 
\begin{equation*}
\begin{tabular}{llllll}
$v_{31}^{\left( 4\right) }=\left( 1,0,\mathrm{\ast }\right) $ & , & $%
v_{32}^{\left( 4\right) }=\left( 0,1,\mathrm{\ast }\right) $ & , & $%
v_{30}^{\left( 4\right) }=\left( -1,-1,\mathrm{\ast }\right) $ & , \\ 
$v_{31}^{\left( 2\right) }=\left( 1,-1,\mathrm{\ast }\right) $ & , & $%
v_{34}^{\left( 2\right) }=\left( 0,-3,\mathrm{\ast }\right) $ & , & $%
v_{30}^{\left( 2\right) }=\left( -1,2,\mathrm{\ast }\right) $ & , \\ 
$v_{32}^{\left( 1\right) }=\left( -1,1,\mathrm{\ast }\right) $ & , & $%
v_{34}^{\left( 1\right) }=\left( -3,0,\mathrm{\ast }\right) $ & , & $%
v_{30}^{\left( 1\right) }=\left( 2,-1,\mathrm{\ast }\right) $ & .%
\end{tabular}%
\end{equation*}

\subsection{Divisors, edges and vertices of toric WP$_{1113}^{3}$}

Like in the case of $P^{3}$, the weighted projective space $WP_{1113}^{3}$
can be recovered by four patches 
\begin{equation}
U_{j}\sim C^{3},\qquad j=1,2,3,4.
\end{equation}%
In the toric language, this weighted projective space is a tetrahedron with
a $T^{3}$ fibration. It has:\newline
\textbf{(i) }four Divisors 
\begin{equation}
F_{1},\text{ \ }F_{2},\text{ \ }F_{3},\text{ \ }F_{4},
\end{equation}%
associated with $w_{i}=0$ on which the torus $T^{3}$ of $WP_{1113}^{3}$\
reduces to $T^{2}$. These divisors are given by 
\begin{equation}
\begin{tabular}{ll}
$F_{1}:|w_{2}|^{2}+|w_{3}|^{2}+3|w_{4}|^{2}=t$ & , \\ 
$F_{2}:|w_{1}|^{2}+|w_{3}|^{2}+3|w_{4}|^{2}=t$ & , \\ 
$F_{3}:|w_{1}|^{2}+|w_{2}|^{2}+3|w_{4}|^{2}=t$ & , \\ 
$F_{4}:|w_{1}|^{2}+|w_{2}|^{2}+|w_{3}|^{2}=t$ & .%
\end{tabular}
\label{re}
\end{equation}%
Three of these faces namely $F_{1},$ $F_{2}$ and $F_{3}$ are isomorphic to $%
WP_{113}^{2}$ with Kahler parameter $t$; but located in different regions of 
$C^{4}$; the fourth is equivalent to $P^{2}$ \newline
\textbf{(ii)} six edges 
\begin{equation}
E_{ij},\qquad i<j,
\end{equation}%
each one given by the intersection of two faces $F_{i}$ and $F_{j}$: 
\begin{equation}
\left\{ 
\begin{array}{c}
E_{12}=F_{1}\cap F_{2},\qquad E_{13}=F_{1}\cap F_{3},\qquad E_{14}=F_{1}\cap
F_{4}, \\ 
E_{23}=F_{2}\cap F_{3},\qquad E_{24}=F_{2}\cap F_{4},\qquad E_{34}=F_{3}\cap
F_{4},%
\end{array}%
\right.
\end{equation}%
where $T^{3}$ of the bulk and $T^{2}$ of the faces shrink down to $S^{1}$. 
\newline
Using the relations(\ref{re}), we can write down the defining eqs of the
geometry associated to these toric edges. We have three projective lines $%
P^{1}$ given by 
\begin{eqnarray}
E_{14} &:&|w_{2}|^{2}+|w_{3}|^{2}=t,  \notag \\
E_{24} &:&|w_{1}|^{2}+|w_{3}|^{2}\ =t, \\
E_{34} &:&|w_{1}|^{2}+|w_{2}|^{2}=t,  \notag
\end{eqnarray}%
and three weighted projective ones $WP_{13}^{1}$ as shown below: 
\begin{eqnarray}
E_{12} &:&|w_{3}|^{2}+3|w_{4}|^{2}=t,  \notag \\
E_{13} &:&|w_{2}|^{2}+3|w_{4}|^{2}=t, \\
E_{23} &:&|w_{1}|^{2}+3|w_{4}|^{2}\ =t.  \notag
\end{eqnarray}%
These lines are located in the different planes of $\mathcal{C}^{4}.$\newline
(\textbf{iii}) four vertices 
\begin{equation}
V_{ijk},\qquad i<j<k,
\end{equation}%
given by the intersection of three faces 
\begin{eqnarray}
V_{123} &=&F_{1}\cap F_{2}\cap F_{3},\qquad V_{124}=F_{1}\cap F_{2}\cap
F_{4},  \notag \\
V_{134} &=&F_{1}\cap F_{3}\cap F_{4},\qquad V_{234}=F_{2}\cap F_{3}\cap
F_{4},
\end{eqnarray}%
where $T^{3}$ of the bulk, the $T^{2}$ of the faces and the $S^{1}$ cycles
of the edges shrink down to zero. These vertices are given by points on the
real lines of $R^{4}\subset $\ $C^{4}$. We have 
\begin{eqnarray}
V_{123} &:&w_{1}=w_{2}=w_{3}=0,\qquad |w_{4}|^{2}=\frac{t}{3},  \notag \\
V_{124} &:&w_{1}=w_{2}=w_{4}=0,\qquad |w_{3}|^{2}=t,  \notag \\
V_{134} &:&w_{1}=w_{3}=w_{4}=0,\qquad |w_{2}|^{2}=t, \\
V_{234} &:&w_{2}=w_{3}=w_{4}=0,\qquad |w_{1}|^{2}=t,  \notag
\end{eqnarray}%
These are just the vertices of a tetraedron. In the next section we show how
this tetraedron leads to build non planar topological formalism for
computing the partition function of the local degenerate elliptic curve in
the large complex structure limit $\mu \rightarrow \infty $.

\section{Non planar topological formalism}

\begin{figure}[tbph]
\begin{center}
\hspace{-1cm} \includegraphics[width=8cm]{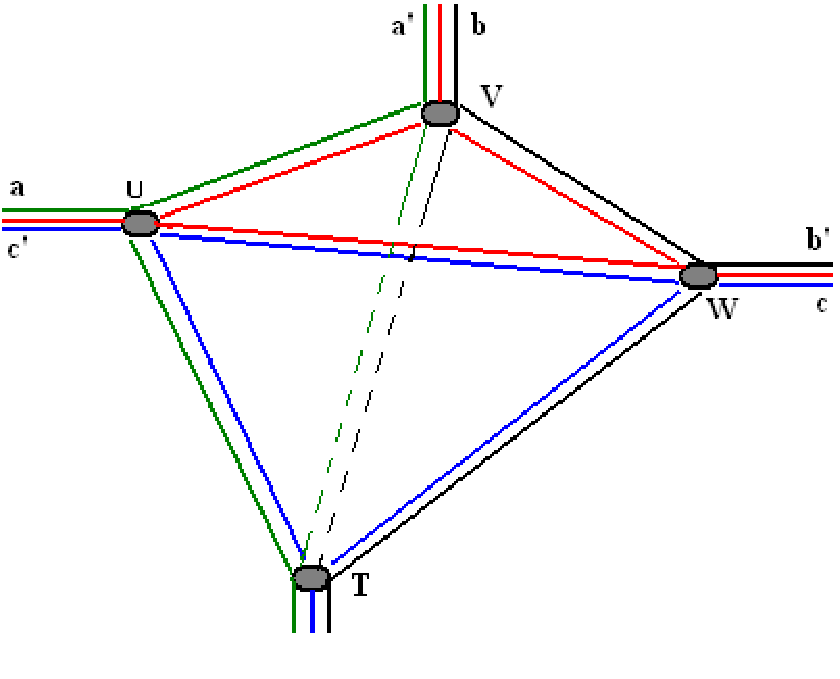}
\end{center}
\par
\vspace{-1cm}
\caption{{\protect\small Toric web-diagram of the local elliptic curve using
planar vertices in different planes. External and internal momenta have been
expressed in terms of 2d partitions.\ }}
\label{mabsy}
\end{figure}

In this section, we consider the example of topological closed string on the
Calabi-Yau threefold hypersurface $H_{3}^{\left( \infty \right) }$. As
remarked earlier, $H_{3}^{\left( \infty \right) }$ is embedded in the normal
bundle of the complex four dimension manifold $\mathcal{O}\left( -3\right)
\rightarrow P^{3}$ with web diagram as in figure (\ref{mabsy}). For this
toric realization, we have the fribration, 
\begin{equation}
H_{3}^{\left( {\small \infty }\right) }=\mathcal{O}\left( -3\right)
\rightarrow S_{2},  \label{bti}
\end{equation}%
where the compact surface $S_{2}$ is roughly $\left[ \partial \left(
P^{3}\right) -\left\{ P^{2}\right\} \right] $ with $\partial \left(
P^{3}\right) $ standing for the boundary of the complex projective space $%
P^{3}$. Notice in passing that $\partial \left( P^{3}\right) $ consists of
four intersecting projective planes; $P_{1}^{2},$ $P_{2}^{2},$ $P_{3}^{2}$
and $P_{4}^{2}$ divisors, as exhibited in the figure (\ref{mabsy}).\newline
\ \newline
Through this particular example, we would like to set up the basis of the
non planar topological vertex formalism for the class of Calabi-Yau
hypersurfaces associated with the supersymmetric gauged sigma model with 
\emph{non zero} superpotential (\ref{w}). More precisely, we consider the
three following things:\newline
(\textbf{1}) we derive the structure of the non planar 3- vertex \QTR{cal}{C}%
$_{3}^{\left( np\right) }$. This vertex can be realized by the combination
of at least two planar topological 3- vertices; say \QTR{cal}{C}$%
_{3}^{\left( xy\right) }$ and \QTR{cal}{C}$_{3}^{\left( yz\right) }$ living
in the $xy$- and $yz$- planes respectively. Note that \QTR{cal}{C}$%
_{3}^{\left( np\right) }$ has an interpretation in terms of \textrm{3d-
partitions }$\Pi ,$ $\Lambda ,$ $\Sigma ,$ and $\Gamma $,\textrm{\ }%
\begin{equation}
\mathcal{C}_{3}^{\left( np\right) }=\mathcal{C}_{\Pi \Lambda \Sigma \Gamma }.
\end{equation}%
(\textbf{2}) we give the explicit expression of the non planar topological
3-vertex \QTR{cal}{C}$_{3}^{\left( np\right) }$; first in terms of products
of the planar 3- vertices and then in terms of products of Schur functions $%
\mathcal{S}_{\xi }$.\newline
(\textbf{3}) we calculate the explicit value of the topological partition
function $\mathcal{Z}_{H_{3}}$ of the Calabi-Yau hypersurface $H_{3}^{\left(
\infty \right) }$ by using \QTR{cal}{C}$_{3}^{\left( np\right) }$. \newline
With this programme in mind, we turn now to give details.

\subsection{Deriving the non planar vertex \QTR{cal}{C}$_{3}^{\left(
np\right) }$}

A direct way to get the structure of the non planar vertex \QTR{cal}{C}$%
_{3}^{\left( np\right) }$ is to start from the web diagram of $H_{3}^{\left(
\infty \right) }$ with the Calabi-Yau condition (\ref{cyc}), 
\begin{equation}
\sum_{i=0}^{4}q_{i}=3.  \label{ccy}
\end{equation}%
Then, use the remarkable relation between the toric graph of $H_{3}^{\left(
\infty \right) }$ and the one corresponding to the normal bundle of the
complex projective plane $NP^{2}$. The compact divisor $S_{2}$ of the
hypersurface $H_{3}^{\left( \infty \right) }$ involves \emph{three}
intersecting complex projective planes $P_{1}^{2}$, $P_{2}^{2}$ and $%
P_{3}^{2}$; i.e 
\begin{equation}
\begin{tabular}{ll}
$S_{2}=P_{1}^{2}\cup P_{2}^{2}\cup P_{3}^{2}$ & ,%
\end{tabular}%
\end{equation}%
with intersections as follows 
\begin{equation}
\begin{tabular}{ll}
$P_{1}^{2}\cap P_{2}^{2}=\mathcal{C}_{3}$ & , \\ 
$P_{2}^{2}\cap P_{3}^{2}=\mathcal{C}_{1}$ & , \\ 
$P_{3}^{2}\cap P_{1}^{2}=\mathcal{C}_{2}$ & ,%
\end{tabular}%
\end{equation}%
where the curves $\mathcal{C}_{1}$, $\mathcal{C}_{2}$ and $\mathcal{C}_{3}$
are projective lines. This property implies that $H_{3}^{\left( \infty
\right) }$ can be obtained by gluing \emph{three} copies of $NP^{2}$ in a
specific manner,

\begin{figure}[tbph]
\begin{center}
\hspace{-1cm} \includegraphics[width=14cm]{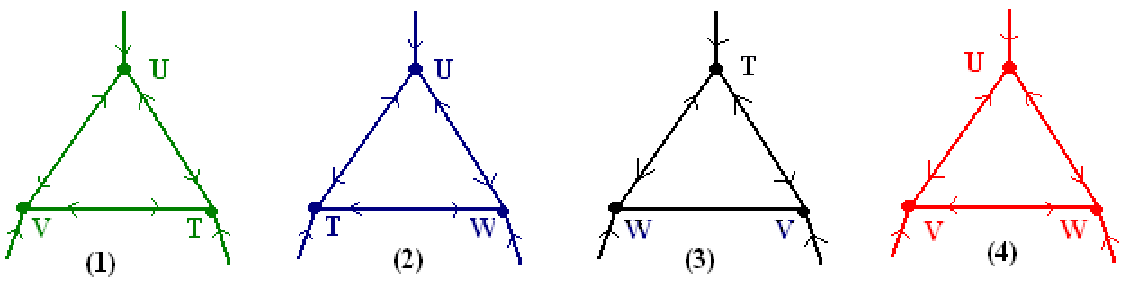}
\end{center}
\caption{{\protect\small These are toric figures presenting the NP}$^{2}$%
{\protect\small \ 's involved in making the CY hypersurface H}$_{3}^{\left(
\infty \right) }${\protect\small . The triangles }$(1)${\protect\small , }$%
(2),$ $(3)$ and $(4)$ belong to different planes of the 3d- space.%
{\protect\small \ The arrows represent the incoming momenta.\ }}
\label{hs}
\end{figure}

\ \ \newline
The various $NP^{2}$'s, represented by the web graphs of the figures (\ref%
{hs}), belong to different planes of the 3d- space and are associated with
the triangles 
\begin{equation}
UVT,\qquad UWT,\qquad VWT.
\end{equation}
To get the hypersurface $H_{3}^{\left( \infty \right) }$ with $T^{2}\times R$
fibration, the triangle $UVW$ in figure (\ref{mabsy}) should be omitted
because of the condition (\ref{ccy}). Notice that like for figure (\ref%
{mabsy}), the web diagram of the hypersurface $H_{3}^{\left( \infty \right)
} $ has four vertices denoted as $U$, $V$, $W$ and $T$ where the respective
incoming momenta $\left\{ \mathbf{u}_{i}\right\} $, $\left\{ \mathbf{v}%
_{i}\right\} $, $\left\{ \mathbf{w}_{i}\right\} $ and $\left\{ \mathbf{t}%
_{i}\right\} $ add to zero, 
\begin{equation}
\begin{tabular}{llllll}
$\sum\limits_{i=1}^{4}\mathbf{u}_{i}=0$ & , & $\sum\limits_{i=1}^{4}\mathbf{v%
}_{i}=0$ & , & $\sum\limits_{i=1}^{4}\mathbf{w}_{i}=0$ & $%
\sum\limits_{i=1}^{4}\mathbf{t}_{i}=0$.%
\end{tabular}%
\end{equation}%
To get more insight on the vectors $\mathbf{u}_{i},$ $\mathbf{v}_{i}$, $%
\mathbf{w}_{i}$ and $\mathbf{t}_{i}$, let us give some details.

\textbf{(i)} \emph{Vertex }$U$\newline
The vector momenta $\mathbf{u}_{1}$, $\mathbf{u}_{2}$, $\mathbf{u}_{3}$ and $%
\mathbf{u}_{4}$, capturing the quantum numbers of the shrinking 1-cycles on
the respective edges $E_{1}$, $E_{2}$, $E_{3}$ and $E_{4}$ ending on the
vertex $U$, are given by the following 3- dimensional integer vectors 
\begin{equation}
U:\left\{ 
\begin{tabular}{lll}
$\mathbf{u}_{1}$ & $=$ & $\left( -1,0,0\right) $ \\ 
$\mathbf{u}_{2}$ & $=$ & $\left( -1,+1,0\right) $ \\ 
$\mathbf{u}_{3}$ & $=$ & $\left( -1,0,+1\right) $ \\ 
$\mathbf{u}_{4}$ & $=$ & $\left( +3,-1,-1\right) $%
\end{tabular}%
\right.
\end{equation}%
The Calabi-Yau condition at the vertex $U$ is given by the following
conservation law, 
\begin{equation}
\mathbf{u}_{1}\mathbf{+u}_{2}\mathbf{+u}_{3}\mathbf{+u}_{4}=\mathbf{0,}
\label{uc}
\end{equation}%
in agreement with the property that all cycles shink to zero at the vertex.
Notice also the two following features:\newline
First, the condition (\ref{uc}) shows that only three of the four vectors $%
\mathbf{u}_{i}$ are linearly independent; they generate a 3d- vector space.%
\newline
Second, the non planar vertex \QTR{cal}{C}$_{3}^{\left( np\right) }$ is
obtained by combining three planar vertices \QTR{cal}{C}$_{3}^{\left(
xy\right) }$, \QTR{cal}{C}$_{3}^{\left( yz\right) }$ and \QTR{cal}{C}$%
_{3}^{\left( zx\right) }$ which come from appropriate projections on 2d-
vector spaces as follows, 
\begin{equation*}
\begin{tabular}{lll}
$U^{xy}:\left\{ 
\begin{tabular}{l}
$\mathbf{u}_{1}^{xy}=\left( -1,0,0\right) $ \\ 
$\mathbf{u}_{2}^{xy}=\left( -1,1,0\right) $ \\ 
$\mathbf{u}_{3}^{\prime xy}=\left( +2,-1,0\right) $%
\end{tabular}%
\right. $, & $U^{yz}:\left\{ 
\begin{tabular}{l}
$\mathbf{u}_{2}^{yz}=\left( 0,1,0\right) $ \\ 
$\mathbf{u}_{3}^{yz}=\left( 0,0,1\right) $ \\ 
$\mathbf{u}_{4}^{yz}=\left( 0,-1,-1\right) $%
\end{tabular}%
\right. $, & $U^{zx}:\left\{ 
\begin{tabular}{l}
$\mathbf{u}_{1}^{zx}=\left( -1,0,0\right) $ \\ 
$\mathbf{u}_{2}^{\prime zx}=\left( +2,0,-1\right) $ \\ 
$\mathbf{u}_{3}^{zx}=\left( -1,0,1\right) $%
\end{tabular}%
\right. $,%
\end{tabular}%
\end{equation*}%
where the upper indices $\left( xy\right) $, $\left( yz\right) $, $\left(
zx\right) $ stand for the $xy$- ,$yz$-, $zx$- planes respectively and where
we have used the folding 
\begin{equation}
\begin{tabular}{lll}
$\mathbf{u}_{3}^{\prime xy}=\mathbf{u}_{3}^{xy}+\mathbf{u}_{4}^{xy}$ & $%
=\left( -1,0,0\right) +\left( 3,-1,0\right) =\left( 2,-1,0\right) $ & , \\ 
$\mathbf{u}_{2}^{\prime zx}=\mathbf{u}_{2}^{zx}+\mathbf{u}_{4}^{zx}$ & $%
=\left( -1,0,0\right) +\left( 3,0,-1\right) =\left( 2,0,-1\right) $ & .%
\end{tabular}%
\end{equation}%
The above 3- vertices $U^{xy}$, $U^{yz}$ and $U^{zx}$, respectively
associated with the topological 3- vertices \QTR{cal}{C}$_{3}^{\left(
xy\right) }$, \QTR{cal}{C}$_{3}^{\left( yz\right) }$ and \QTR{cal}{C}$%
_{3}^{\left( zx\right) }$, obey the Calabi-Yau conditions 
\begin{equation}
\begin{tabular}{ll}
$\mathbf{u}_{1}^{xy}+\mathbf{u}_{2}^{xy}+\mathbf{u}_{3}^{\prime xy}=\mathbf{0%
}$ & , \\ 
$\mathbf{u}_{2}^{yz}+\mathbf{u}_{3}^{yz}+\mathbf{u}_{4}^{yz}=\mathbf{0}$ & ,
\\ 
$\mathbf{u}_{1}^{zy}+\mathbf{u}_{2}^{\prime zy}+\mathbf{u}_{3}^{zy}=\mathbf{0%
}$ & .%
\end{tabular}%
\end{equation}%
As a result, the non planar topological vertex \QTR{cal}{C}$_{3}^{\left(
np\right) }$ is built from the 3- vertices \QTR{cal}{C}$_{3}^{\left(
xy\right) }$, \QTR{cal}{C}$_{3}^{\left( yz\right) }$ and \QTR{cal}{C}$%
_{3}^{\left( zx\right) }$ associated with the patches $U^{xy}$, $U^{yz}$ and 
$U^{zx}$. Notice that since $U^{xy}$, $U^{yz}$ and $U^{zx}$ belong to
different planes, \QTR{cal}{C}$_{3}^{\left( np\right) }$ is then a non
planar vertex. \newline

\textbf{(ii)} \emph{Vertex V}\newline
The incoming momenta $\left\{ \mathbf{v}_{1},\mathbf{v}_{2},\mathbf{v}_{3},%
\mathbf{v}_{4}\right\} $ ending at the vertex $V$ of the web diagram of $%
H_{3}^{\left( \infty \right) }$ read as follows 
\begin{equation}
V:\left\{ 
\begin{tabular}{lll}
$\mathbf{v}_{1}$ & $=$ & $\left( -1,-1,+3\right) $ \\ 
$\mathbf{v}_{2}$ & $=$ & $\left( +1,0,-1\right) $ \\ 
$\mathbf{v}_{3}$ & $=$ & $\left( 0,+1,-1\right) $ \\ 
$\mathbf{v}_{4}$ & $=$ & $\left( 0,0,-1\right) $%
\end{tabular}%
\ \right. .
\end{equation}%
The planar patches making the non planar topological vertex $V$ are given
by, 
\begin{equation*}
\begin{tabular}{lll}
$V^{xy}:\left\{ 
\begin{tabular}{l}
$\mathbf{v}_{1}^{xy}=\left( {\small -1,-1,0}\right) $ \\ 
$\mathbf{v}_{2}^{xy}=\left( {\small +1,0,0}\right) $ \\ 
$\mathbf{v}_{3}^{xy}=\left( {\small 0,+1,0}\right) $%
\end{tabular}%
\right. $, & $V^{yz}:\left\{ 
\begin{tabular}{l}
$\mathbf{v}_{1}^{\prime yz}=\left( {\small 0,-1,+2}\right) $ \\ 
$\mathbf{v}_{2}^{yz}=\left( {\small 0,0,-1}\right) $ \\ 
$\mathbf{v}_{3}^{yz}=\left( {\small 0,+1,-1}\right) $%
\end{tabular}%
\right. $, & $V^{zx}:\left\{ 
\begin{tabular}{l}
$\mathbf{v}_{1}^{\prime zx}=\left( {\small -1,0,+2}\right) $ \\ 
$\mathbf{v}_{2}^{zx}=\left( {\small +1,0,-}1\right) $ \\ 
$\mathbf{v}_{3}^{zx}=\left( {\small 0,0,-1}\right) $%
\end{tabular}%
\right. $,%
\end{tabular}%
\end{equation*}%
They satisfy the conservation laws 
\begin{equation}
\begin{tabular}{ll}
$\mathbf{v}_{1}^{xy}+\mathbf{v}_{2}^{xy}+\mathbf{v}_{3}^{xy}=\mathbf{0}$ & ,
\\ 
$\mathbf{v}_{1}^{\prime yz}+\mathbf{v}_{2}^{yz}+\mathbf{v}_{3}^{yz}=\mathbf{0%
}$ & , \\ 
$\mathbf{v}_{1}^{\prime zy}+\mathbf{v}_{2}^{zy}+\mathbf{v}_{3}^{zy}=\mathbf{0%
}$ & ,%
\end{tabular}%
\end{equation}%
encoding the properties that all cycles shrink at the vertex $V$.

\textbf{(iii)}\emph{\ Vertex W}\newline
Similarly, the incoming momenta $\left\{ \mathbf{w}_{1},\mathbf{w}_{2},%
\mathbf{w}_{3},\mathbf{w}_{4}\right\} $ at the vertex $W$ read as follows, 
\begin{equation}
W:\left\{ 
\begin{array}{c}
\mathbf{w}_{1}=\left( {\small -1,3,-1}\right) , \\ 
\mathbf{w}_{2}=\left( +{\small 1,-1,0}\right) , \\ 
\mathbf{w}_{3}=\left( {\small 0,-1,+1}\right) , \\ 
\mathbf{w}_{4}=\left( {\small 0,-1,0}\right) .%
\end{array}%
\right.
\end{equation}%
We also have the following planar patches 
\begin{equation*}
\begin{tabular}{lll}
$W^{xy}:\left\{ 
\begin{array}{c}
\mathbf{w}_{1}^{\prime xy}\mathbf{=}\left( -1,2,0\right) \\ 
\mathbf{w}_{2}^{xy}\mathbf{=}\left( 1,-1,0\right) \\ 
\mathbf{w}_{3}^{xy}\mathbf{=}\left( 0,-1,0\right)%
\end{array}%
\right. $, & $W^{yz}:\left\{ 
\begin{array}{c}
\mathbf{w}_{1}^{\prime yz}\mathbf{=}\left( 0,2,-1\right) \\ 
\mathbf{w}_{2}^{yz}\mathbf{=}\left( 0,-1,0\right) \\ 
\mathbf{w}_{3}^{yz}\mathbf{=}\left( 0,-1,1\right)%
\end{array}%
\right. $, & $W^{zx}:\left\{ 
\begin{array}{c}
w_{1}^{zx}=\left( -1,0,-1\right) \\ 
w_{2}^{zx}=\left( 1,0,0\right) \\ 
w_{3}^{zx}=\left( 0,0,1\right)%
\end{array}%
\right. $,%
\end{tabular}%
\end{equation*}%
The momenta satisfy the Calabi-Yau conditions: 
\begin{equation}
\begin{tabular}{ll}
$\mathbf{w}_{1}^{\prime xy}+\mathbf{w}_{2}^{xy}+\mathbf{w}_{3}^{xy}=\mathbf{0%
}$ & , \\ 
$\mathbf{w}_{1}^{\prime yz}+\mathbf{w}_{2}^{yz}+\mathbf{w}_{3}^{yz}=\mathbf{0%
}$ & , \\ 
$\mathbf{w}_{1}^{zy}+\mathbf{w}_{2}^{zy}+\mathbf{w}_{3}^{zy}=\mathbf{0}$ & .%
\end{tabular}%
\end{equation}

\textbf{(iv)}\emph{\ Vertex T}\newline
This vertex, given by 
\begin{equation}
T:\left\{ 
\begin{tabular}{lll}
$\mathbf{t}_{1}$ & $=$ & $\left( -1,-1,-1\right) $ \\ 
$\mathbf{t}_{2}$ & $=$ & $\left( 1,0,0\right) $ \\ 
$\mathbf{t}_{3}$ & $=$ & $\left( 0,1,0\right) $ \\ 
$\mathbf{t}_{4}$ & $=$ & $\left( 0,0,1\right) $%
\end{tabular}%
\ \right. ,
\end{equation}%
is, in some sense a special vertex since it has a leg related to an external
source with non zero momentum $\mathbf{t}_{1}$. The planar patches $T^{xy},$ 
$T^{yz}$\ and $T^{zx}$\ forming the non planar topological vertex $T$ are
given by 
\begin{equation*}
\begin{tabular}{lll}
$T^{xy}:\left\{ 
\begin{tabular}{l}
$\mathbf{t}_{1}^{xy}=\left( -1,-1,0\right) $ \\ 
$\mathbf{t}_{2}^{xy}=\left( 1,0,0\right) $ \\ 
$\mathbf{t}_{3}^{xy}=\left( 0,1,0\right) $%
\end{tabular}%
\right. $, & $T^{yz}:\left\{ 
\begin{tabular}{l}
$\mathbf{t}_{1}^{yz}=\left( 0,-1,-1\right) $ \\ 
$\mathbf{t}_{3}^{yz}=\left( 0,1,0\right) $ \\ 
$\mathbf{t}_{4}^{yz}=\left( 0,0,1\right) $%
\end{tabular}%
\right. $, & $T^{zx}:\left\{ 
\begin{tabular}{l}
$\mathbf{t}_{1}^{zx}=\left( -1,0,-1\right) $ \\ 
$\mathbf{t}_{2}^{zx}=\left( 1,0,0\right) $ \\ 
$\mathbf{t}_{4}^{zx}=\left( 0,0,1\right) $%
\end{tabular}%
\right. $,%
\end{tabular}%
\end{equation*}%
with the Calabi-Yau conditions 
\begin{equation}
\begin{tabular}{ll}
$\mathbf{t}_{1}^{xy}+\mathbf{t}_{2}^{xy}+\mathbf{t}_{3}^{xy}=\mathbf{0}$ & ,
\\ 
$\mathbf{t}_{1}^{yz}+\mathbf{t}_{3}^{yz}+\mathbf{t}_{4}^{yz}=\mathbf{0}$ & ,
\\ 
$\mathbf{t}_{1}^{zx}+\mathbf{t}_{2}^{zx}+\mathbf{t}_{4}^{zx}=\mathbf{0}$ & .%
\end{tabular}%
\end{equation}%
Having derived the structure of the non planar vertex that is involved in
the Calabi-Yau hypersurface $H_{3}^{\left( \infty \right) }$, we turn now to
compute its expression in terms of the usual planar ones.

\subsection{Non planar topological vertex \QTR{cal}{C}$_{3}^{\left(
np\right) }$}

To get the explicit expression of the non planar topological vertex 
\QTR{cal}{C}$_{3}^{\left( np\right) }$, we first focus on the vertex $U$.
Then, we extend the obtained results to the other vertices $V$, $W$ and $T$. 
\newline
A priori, a generic non planar vertex is made of at least two planar
vertices belonging to different planes as shown in the figure (\ref{mabsy}).
Based on this observation and motivated by the works \textrm{\cite{33,31}},
we deduce\textrm{\footnote{%
More details are presented in the appendix.}} that the expression of 
\QTR{cal}{C}$_{3}^{\left( np\right) }$, associated with the vertex $U$, is
given by the following relation 
\begin{equation}
\mathcal{C}_{3}^{\left( np\right) }=\mathcal{C}_{3}^{\left( xy\right)
}\times \mathcal{C}_{3}^{\left( zx\right) },  \label{cc}
\end{equation}%
where the expression of \QTR{cal}{C}$_{3}^{\left( xy\right) }$ and \QTR{cal}{%
C}$_{3}^{\left( zx\right) }$ are topological 3-vertices as computed in 
\textrm{\cite{vafa}}. To get the expression of \QTR{cal}{C}$_{3}^{\left(
np\right) }$ in terms of Schur functions, let us recall some useful results
on the planar 3-vertex formalism.

\emph{Planar} \emph{topological 3- vertex formalism} \newline
First recall that a toric Calabi-Yau threefold with a $T^{2}\times R$
special Lagrangian fibration has toric geometry represented by planar web
diagrams. Following \textrm{\cite{vafa}}, the expression of the topological
3-vertex $\mathcal{C}_{\emptyset \emptyset \emptyset }$ without boundary
conditions; i.e $\left( \lambda ,\mu ,\nu \right) =$ $\left( \emptyset
,\emptyset ,\emptyset \right) $, is given by\footnote{%
The usual topological vertex $C_{\lambda \mu \nu }$ considered in \cite{vafa}
is planar. In our study, it should be thought of either as $C_{_{\lambda \mu
\nu }}^{xy}$ or $C_{_{\lambda \mu \nu }}^{yz}$ or again as $C_{_{\lambda \mu
\nu }}^{zx}.$} 
\begin{equation}
\mathcal{C}_{\emptyset \emptyset \emptyset }\left( q\right)
=\dprod\limits_{k=1}^{\infty }\left( \frac{1}{\left( 1-q^{k}\right) ^{k}}%
\right) ,
\end{equation}%
and describes the topological closed string amplitude on $C^{3}$ with $%
q=e^{-g_{s}}$\ and $g_{s}$\ being the closed string coupling constant. It
happens that this relation is nothing but the 3d- MacMahon function $%
\mathcal{Z}_{3d}$ generating plane partitions. \newline
Open strings ending on D- branes are implemented by introducing non trivial
boundary conditions $\left( \lambda ,\mu ,\nu \right) $ on the edges of the
3-vertex. The topological 3- vertex associated with this configuration is
denoted as $\mathcal{C}_{\lambda \mu \nu }(q)$ and its contribution reads,
in terms of the Schur functions $\mathcal{S}_{\xi }\left( q\right) $, as
follows 
\begin{equation}
\mathcal{C}_{\lambda \mu \nu }(q)=q^{\kappa (\lambda )}\left[ \mathcal{S}%
_{\nu ^{T}}(q^{-\rho })\sum_{2d\text{ partitions }\eta }\mathcal{S}_{\lambda
^{T}/\eta }(q^{-\nu -\rho })\mathcal{S}_{\mu /\eta }(q^{-\nu ^{T}-\rho })%
\right] .  \label{cd}
\end{equation}%
where $\kappa (\lambda )=2\left( \left\Vert \lambda \right\Vert
^{2}-\left\vert \lambda \right\vert \right) -2\left( \left\Vert \lambda
^{T}\right\Vert ^{2}-\left\vert \lambda ^{T}\right\vert \right) $ is related
to the Casimir of the representation of 2d partition $\lambda $, while $\nu
^{T}$ represents the transpose of the Young diagram and $\mathcal{S}%
_{\lambda /\eta }(q)$ stands for the skew Schur function with $q^{-\nu -\rho
}=\left( q^{-\nu _{1}-\frac{1}{2}},q^{-\nu _{2}-\frac{3}{2}},q^{-\nu _{3}-%
\frac{5}{2}}\ldots \right) $.

\emph{Determining} \QTR{cal}{C}$_{3}^{\left( np\right) }$\newline
At $U$, the non planar topological vertex \QTR{cal}{C}$_{3}^{\left(
np\right) }$ (\ref{cc}) reads, by implementing the boundary conditions, as
follows, 
\begin{equation}
\mathcal{C}_{\mathrm{abcdef}}^{\left( np\right) }\left( q_{1},q_{2}\right) =%
\mathcal{C}_{\mathrm{ab}\text{\textrm{c}}}^{\left( xy\right) }\left(
q_{1}\right) \times \delta _{\text{\textrm{c}}d}\times \mathcal{C}_{\mathrm{%
def}}^{\left( zx\right) }\left( q_{2}\right) ,  \label{np1}
\end{equation}%
where $q_{1}$ and $q_{2}$ are two parameters which may be set as $%
q_{1}=q_{2}=e^{-g_{s}}$. The a, b, c, d, e, f\ stand for 2d- partitions
encoding the configuration of D- branes which end on each edge of the toric
diagram. The 3- vertex $\mathcal{C}_{\mathrm{abc}}^{\left( xy\right) }$ is a
planar vertex in $xy$- plane while $\mathcal{C}_{\mathrm{def}}^{\left(
zx\right) }$ is planar in the $zx$- plane. The factor $\delta _{\text{%
\textrm{c}}d}$ of eq(\ref{np1}) captures the data on the intersection
between $\mathcal{C}_{\mathrm{abc}}^{\left( xy\right) }$ and $\mathcal{C}_{%
\mathrm{def}}^{\left( zx\right) }$. Using eq(\ref{cd}), we get 
\begin{equation*}
\begin{tabular}{lll}
\QTR{cal}{C}$_{\mathrm{abcdef}}^{\left( np\right) }\left( q_{1},q_{2}\right)
=$ & $\delta _{\text{\textrm{c}}d}$ $q_{1}^{\kappa (\mathrm{a})}\left[ 
\mathcal{S}_{\text{\textrm{c}}^{T}}(q_{1}^{-\rho })\sum_{2d\text{ partitions 
}\eta }\mathcal{S}_{\mathrm{a}^{T}/\eta }(q_{1}^{-\text{\textrm{c}}-\rho })%
\mathcal{S}_{\mathrm{b}/\eta }(q_{1}^{-\text{\textrm{c}}^{T}-\rho })\right] $
&  \\ 
& $\times $ $q_{2}^{\kappa (\mathrm{d})}\left[ \mathcal{S}_{\mathrm{f}%
^{T}}(q_{2}^{-\rho })\sum_{2d\text{ partitions }\xi }\mathcal{S}_{\mathrm{d}%
^{T}/\xi }(q_{2}^{-\mathrm{f}-\rho })\mathcal{S}_{\mathrm{e}/\xi }(q_{2}^{-%
\mathrm{f}^{T}-\rho })\right] $ & .%
\end{tabular}%
\end{equation*}%
Similar relations are valid for the vertices $V$ and $W$; they read as
follows, 
\begin{equation}
\begin{tabular}{llll}
\QTR{cal}{C}$_{\mathrm{a}^{\prime }\mathrm{b}^{\prime }\text{\textrm{c}}%
^{\prime }\text{d}^{\prime }\mathrm{e}^{\prime }\mathrm{f}^{\prime
}}^{\left( np\right) }\left( V\right) $ & $=$ & \QTR{cal}{C}$_{\mathrm{a}%
^{\prime }\mathrm{b}^{\prime }\text{\textrm{c}}^{\prime }\text{d}^{\prime }%
\mathrm{e}^{\prime }\mathrm{f}^{\prime }}^{\left( np\right) }\left(
q_{2},q_{3}\right) $ & , \\ 
\QTR{cal}{C}$_{\mathrm{a}"\mathrm{b"c"d}^{\prime \prime }\mathrm{e"f"}%
}^{\left( np\right) }\left( W\right) $ & $=$ & \QTR{cal}{C}$_{\mathrm{%
a"b"c"d"e"f"}}^{\left( np\right) }\left( q_{3},q_{1}\right) $ & .%
\end{tabular}%
\end{equation}%
Regarding the patch $T$, the corresponding non planar topological vertex 
\begin{equation*}
\mathcal{C}_{\mathrm{\alpha \beta \gamma \delta \zeta \theta \lambda \mu \nu 
}}^{\left( np\right) }\left( T\right)
\end{equation*}%
is made of three planar topological 3-vertices $\mathcal{C}_{\mathrm{\alpha
\beta \gamma }}\left( T^{zx}\right) $, $\mathcal{C}_{\mathrm{\delta \zeta
\theta }}\left( T^{yz}\right) $ and $\mathcal{C}_{\mathrm{\lambda \mu \nu }%
}\left( T^{zx}\right) $. We have 
\begin{equation}
\mathcal{C}_{\mathrm{\alpha \beta \gamma \delta \zeta \theta \lambda \mu \nu 
}}^{\left( np\right) }\left( T\right) =\mathcal{C}_{\mathrm{\alpha \beta
\gamma }}^{\left( xy\right) }\left( q_{1}\right) \times \delta _{\gamma
\delta }\times \mathcal{C}_{\mathrm{\delta \zeta \theta }}^{\left( yz\right)
}\left( q_{2}\right) \times \delta _{\mathrm{\theta \lambda }}\times 
\mathcal{C}_{\mathrm{\lambda \mu \nu }}^{\left( zx\right) }\left(
q_{3}\right) \times \delta _{\beta \nu },  \label{np2}
\end{equation}%
where $\mathcal{C}_{\mathrm{\alpha \beta \gamma }}^{\left( xy\right) }\left(
q_{1}\right) $, $\mathcal{C}_{\mathrm{\delta \zeta \theta }}^{\left(
yz\right) }\left( q_{2}\right) $ and $\mathcal{C}_{\mathrm{\lambda \mu \nu }%
}^{\left( zx\right) }\left( q_{3}\right) $ are as in eq(\ref{cd}).

\subsection{Explicit expression of $\mathcal{Z}_{H_{3}}$}

\begin{figure}[tbph]
\begin{center}
\hspace{-1cm} \includegraphics[width=10cm]{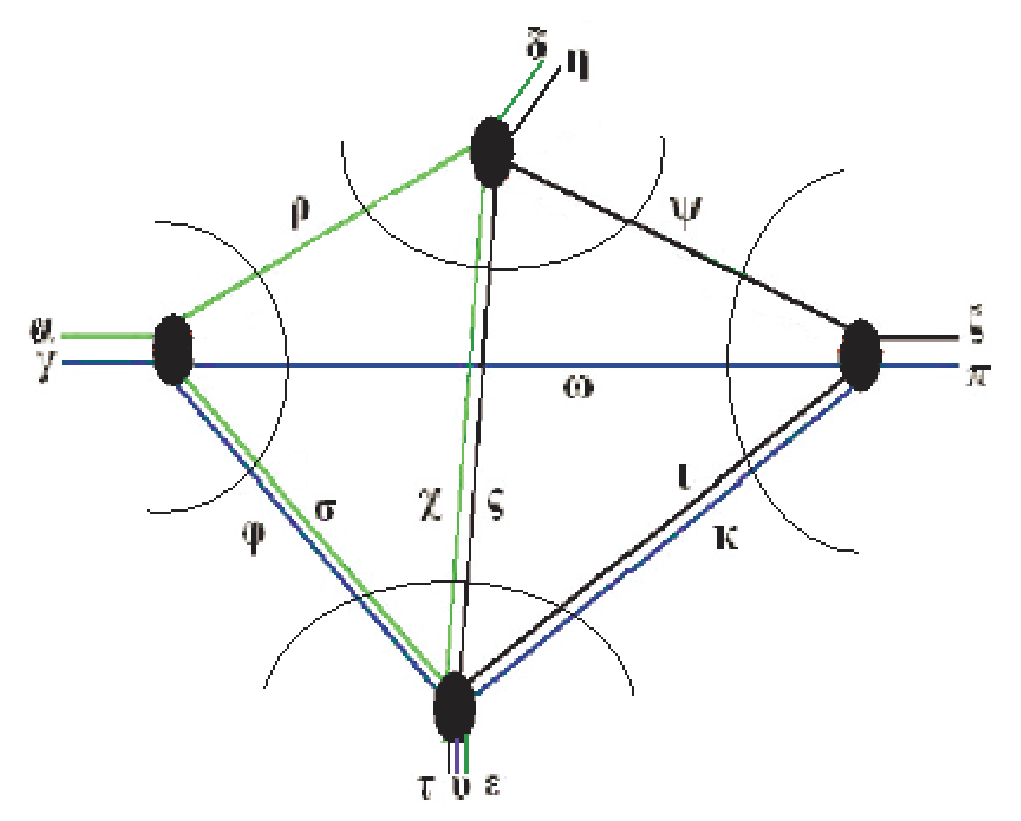}
\end{center}
\par
\vspace{-1cm}
\caption{{\protect\small Non planar graph representing elliptic curve as a
toric manifold. Cutting the graph into four non planar 3-vertex.}}
\label{pfp3}
\end{figure}

In this subsection, we derive the expression of $\mathcal{Z}_{H_{3}}$ by
using the non planar topological vertex formalism. Starting from the diagram
(\ref{mabsy}), which we put it in the form (\ref{pfp3}), we can determine $%
\mathcal{Z}_{H_{3}}$ by help of the cutting and gluing method of \emph{%
Aganagic et al}.\newline
\ \newline
Mimicking the planar topological vertex formalism and using eq(\ref{np1}-\ref%
{np2}), we obtain 
\begin{equation}
\mathcal{Z}_{X_{3}}=\sum_{\left\{ \varkappa \right\} }\left[ \left( \mathcal{%
U}_{\omega \varphi ^{\mathrm{T}}\rho \sigma ^{\mathrm{T}}}\right) \left( 
\mathcal{V}_{\chi \rho ^{\mathrm{T}}\psi \varsigma ^{\mathrm{T}}}\right)
\left( \mathcal{W}_{\iota \psi ^{\mathrm{T}}\kappa \omega ^{\mathrm{T}%
}}\right) \left( \mathcal{T}_{\varphi \kappa ^{\mathrm{T}}\chi \sigma ^{%
\mathrm{T}}\iota \varsigma ^{\mathrm{T}}}\right) \mathcal{H}_{\omega \varphi
\rho \sigma \chi \iota \kappa \psi \varsigma }\right]
\end{equation}%
where $\varkappa $ stands for the 2d- partitions $\omega ,$ $\varphi ,$ $%
\rho ,$ $\sigma ,$ $\chi ,$ $\psi ,$ $\varsigma ,$ $\iota ,$ $\kappa $ while 
$\mathcal{U}_{\omega \varphi ^{\mathrm{T}}\rho \sigma ^{\mathrm{T}}}$, $%
\mathcal{V}_{\chi \rho ^{\mathrm{T}}\psi \varsigma ^{\mathrm{T}}}$, $%
\mathcal{W}_{\iota \psi ^{\mathrm{T}}\kappa \omega ^{\mathrm{T}}}$, and $%
\mathcal{T}_{\varphi \kappa ^{\mathrm{T}}\chi \sigma ^{\mathrm{T}}\iota
\varsigma ^{\mathrm{T}}}$ \ represent the non planar vertex\ which read in
terms of the planar topological 3-vertex $\mathcal{C}_{\alpha \beta \gamma }$
as follows 
\begin{equation}
\begin{tabular}{llllll}
$\mathcal{U}_{\omega \varphi ^{\mathrm{T}}\rho \sigma ^{\mathrm{T}}}$ & $=$
& \QTR{cal}{C}$_{\emptyset \omega \varphi ^{\mathrm{T}}\emptyset \rho \sigma
^{\mathrm{T}}}^{\left( np\right) }\left( q_{1},q_{2}\right) $ & $=$ & $%
\delta _{\varphi ^{\mathrm{T}}\sigma ^{\mathrm{T}}}\times \mathcal{C}%
_{\emptyset \omega \varphi ^{\mathrm{T}}}(q_{1})\mathcal{C}_{\emptyset \rho
\sigma ^{\mathrm{T}}}\left( q_{2}\right) $ & $,$ \\ 
$\mathcal{V}_{\chi \rho ^{\mathrm{T}}\psi \varsigma ^{\mathrm{T}}}$ & $=$ & 
\QTR{cal}{C}$_{\emptyset \chi \rho ^{\mathrm{T}}\emptyset \psi \varsigma ^{%
\mathrm{T}}}^{\left( np\right) }\left( q_{2},q_{3}\right) $ & $=$ & $\delta
_{\chi \varsigma ^{\mathrm{T}}}\times \mathcal{C}_{\emptyset \chi \rho ^{%
\mathrm{T}}}(q_{2})\mathcal{C}_{\emptyset \psi \varsigma ^{\mathrm{T}%
}}\left( q_{3}\right) $ & $,$ \\ 
$\mathcal{W}_{\iota \psi ^{\mathrm{T}}\kappa \omega ^{\mathrm{T}}}$ & $=$ & 
\QTR{cal}{C}$_{\emptyset \iota \psi ^{\mathrm{T}}\emptyset \kappa \omega ^{%
\mathrm{T}}}^{\left( np\right) }\left( q_{3},q_{1}\right) $ & $=$ & $\delta
_{\iota \kappa }\times \mathcal{C}_{\emptyset \iota \psi ^{\mathrm{T}%
}}(q_{3})\mathcal{C}_{\emptyset \kappa \omega ^{\mathrm{T}}}\left(
q_{1}\right) $ & $,$%
\end{tabular}%
\end{equation}%
and%
\begin{equation}
\begin{tabular}{llll}
$\mathcal{T}_{\sigma \chi ^{\mathrm{T}}\iota ^{\mathrm{T}}\varphi \kappa ^{%
\mathrm{T}}\varsigma ^{\mathrm{T}}}$ & $=$ & \QTR{cal}{C}$_{\sigma \chi ^{%
\mathrm{T}}\iota ^{\mathrm{T}}\emptyset \kappa \omega ^{\mathrm{T}}}^{\left(
np\right) }\left( q_{3},q_{1}\right) $ & , \\ 
& $=$ & $\delta _{\iota ^{\mathrm{T}}\kappa ^{\mathrm{T}}}\mathcal{C}%
_{\sigma \chi ^{\mathrm{T}}\emptyset }\left( q_{1}\right) \delta _{\sigma
\varphi }\mathcal{C}_{\varphi \kappa ^{\mathrm{T}}\emptyset }\left(
q_{2}\right) \delta _{\chi ^{\mathrm{T}}\varsigma }\mathcal{C}_{\varsigma
\iota ^{\mathrm{T}}\emptyset }\left( q_{3}\right) $ & .%
\end{tabular}%
\end{equation}%
The term $\mathcal{H}_{\omega \varphi \rho \sigma \chi \psi \varsigma \iota
\kappa }=\prod\nolimits_{\varkappa =\omega ,\varphi ,\rho ,\sigma ,\chi
,\psi ,\varsigma ,\iota ,\kappa }(-e^{-t})^{\left\vert \varkappa \right\vert
}q^{\kappa \left( \varkappa \right) }$ is the framing factor.

\section{Conlusion}

In this paper, we have studied the topological string theory on the special
class of toric Calabi-Yau hypersurfaces which are realized in terms of
supersymmetric gauged linear sigma model with non zero gauge invariant
superpotential $\mathcal{W}\left( \Phi \right) \neq 0$. \newline
Recall that for the case where there is no matter self- interaction ($%
\mathcal{W}\left( \Phi \right) =0$) the Calabi-Yau threefold is described by
the equations of motion (\ref{da}) of the auxiliary fields D$^{a}$ in the
gauge multiplets. In this case the topological 3- vertex is planar and the
topological string amplitudes on this kind of toric Calabi-Yau threefolds
with $T^{2}\times R$ fibration, is obtained by cutting and gluing method as
done in \textrm{\cite{vafa}}. \newline
In the case where $\mathcal{W}\left( \Phi \right) \neq 0$, one has moreover
extra constraint eqs on the complex scalar field variables coming from the
equations of motion of the auxiliary fields F in the chiral superfields. The
resulting local Calabi-Yau threefolds are still toric; but the topological
vertex is non planar.\newline
In the present study we have made a step towards the developments of non
planar topological vertex formalism by focusing on the example of the
Calabi-Yau hypersurface $H_{3}$ with $R\times T^{2}$ fibration. We have
derived the general structure of the non planar topological vertex \QTR{cal}{%
C}$_{3}^{\left( np\right) }$ and its explicit expression as a product of the
planar ones. We have also used this formalism to compute the partition
function of $H_{3}$. From the analysis on the example of the local elliptic
curve in large complex structure, we have learnt \ that the non planar
vertex shares features with the 4- vertex of CY4- folds and has an
interpretation in terms of 3d- partitions. Further progress in this issues
will be given in a future occasion.

\begin{acknowledgement}
\qquad {\small \ \ \ }\newline
{\small This research work is supported by the program Protars III D12/25.
BD and HJ would like to thank ICTP for kind hospitality where part of this
work has been done.}
\end{acknowledgement}

\section{Appendix}

For solving the local Gromov-Witten theory of curves, several methods have
been developed. One of them has been worked out by Bryan and Pandharipande
in \textrm{\cite{31,32}} by using the localization and degeneration methods.
The basic integrals in the local Gromov-Witten theory of $P^{1}$ are
evaluated exactly by localization while the degeneration is required to
capture higher genus curves. Amongst the interesting results gotten in 
\textrm{\cite{31,32}}; we quote the computation of the partition function of%
\textrm{\ }local Gromov-Witten invariants of curves in Calabi Yau
threefolds. The degeneration\textrm{\ }method corresponds to splitting a
genus $g$ surface $X$ along a separating non-singular divisor $B\subset X$
to obtain two surfaces $(X_{1},B)$ and $(X_{2},B)$ of genus $g_{1}$ and $%
g_{2}$.

\begin{equation}
\begin{tabular}{lll}
$Z(X)$ & $=$ & $\sum\limits_{\lambda }Z(X_{1})^{\lambda }Z(X_{2})_{\lambda
}. $%
\end{tabular}
\label{gre}
\end{equation}%
It follows from this analysis that local Gromov-Witten theory of curves is
closely related to q-deformed 2D Yang-Mills theory and bound states of BPS
black holes obtained by the string theoretic method \textrm{\cite{t3}}. More
explicitly; Vafa used the topological vertex method for a particular class
of local threefolds involving the total space of a direct sum of a line
bundle and its inverse on elliptic curve $T^{2}$. This study leads to the
following partition function of $2d$ $U(N)$ Yang-Mills on $T^{2}$%
\begin{equation*}
Z^{YM}\sim Z_{+}^{YM}Z_{-}^{YM}=Z_{top}\bar{Z}_{top}
\end{equation*}%
that appears as a product of a holomorphic and an anti-holomorphic partition
function and where $Z_{top}$ is expressed as follows%
\begin{equation*}
Z_{top}=\sum\limits_{R}q^{mk(R)/2}exp(-t|R|).
\end{equation*}%
Recall that $k(R)$ is the term of Casimir, $|R|$ the number of boxes of the
Young diagram R, $exp(-t|R|)$ is the propagator and $q^{mk(R)/2}$.is the
framing factor. This partition function\textrm{\ }coincides exactly with the
one calculed in page 34 of \textrm{\cite{31}}.\newline
In our present work, we have presented the main lines of the non-planar
topological vertex formalism to solving the theory of local curves for the
special class of toric Calabi-Yau hypersurfaces. The formalism used here for
computing the expression of partition function of degenerate elliptic curve $%
g=1$ can also be extended to compute the partition function of degenerate
higher genus elliptic curve. For the case $g=2$ for instance, the closed
topological string partition function of the local $g$- Riemann surface in
the large complex structures limit, we have 
\begin{equation*}
Z\left( q,Q\right) =\sum_{\xi _{1},\xi _{2}}\left( -Q\right) ^{\left\vert
\xi _{1}\right\vert +\left\vert \xi _{2}\right\vert }Z_{\xi _{1},\xi
_{2}}^{\left( g=1\right) }(q)f_{\xi _{1},\xi _{2}}Z_{\xi _{1}^{T},\xi
_{2}^{T}}^{\left( g=1\right) }
\end{equation*}%
and is obtained by gluing two genus $g=1$ open string partition functions
(5.24) together to form the closed string partition function.\newline
Notice that the non planar topological vertex eq(5.24), deduced from the
ramification of the non planar vertex as the union of two planar topological
vertices $C^{1}\cup C^{2}$ in the two distinct plane $xy$ and $zx$
respectively eq(5.18), has a connection with the mathematical and stringy
methods. presented previously. This link follows obviously from the
identification\ of eq(5.18) with Bryan and Pandharipande formula (\ref{gre}).%
\newline
From the above brief description; it follows that our non-planar topological
vertex formalism should be though of as a different, but equivalent, way for
solving the theory of local curves.

\end{document}